\documentclass[11pt,a4paper]{article}
\usepackage{amsfonts}
\usepackage{amsmath}
\usepackage{amssymb}
\usepackage{jheppub}
\usepackage{ulem}
\usepackage{extarrows}
\usepackage{multirow}
\usepackage{float}

\usepackage{inputenc}
\usepackage{xspace}
\usepackage{url}
\usepackage{epstopdf}

 \def\be{\begin{equation}}
\def\ee{\end{equation}}
\def\bea{\begin{eqnarray}}
\def\eea{\end{eqnarray}}

\def\a{\alpha}
\def\b{\beta}
\def\g{\gamma}

\def\s{\sigma}

\def\bp{\bar{p}}

 \def\p{\partial}

 \def\IZ{\relax\ifmmode\mathchoice
 {\hbox{\cmss Z\kern-.4em Z}}{\hbox{\cmss Z\kern-.4em Z}}
 {\lower.9pt\hbox{\cmsss Z\kern-.4em Z}}
 {\lower1.2pt\hbox{\cmsss Z\kern-.4em Z}}\else{\cmss Z\kern-.4em Z}\fi}
 \def\IB{\relax{\rm I\kern-.18em B}}
 \def\IC{{\relax\hbox{$\inbar\kern-.3em{\rm C}$}}}
 \def\Ic{{\relax\hbox{$\inbar\kern-.22em{\rm c}$}}}
 \def\ID{\relax{\rm I\kern-.18em D}}
 \def\IE{\relax{\rm I\kern-.18em E}}
 \def\IF{\relax{\rm I\kern-.18em F}}
 \def\IG{\relax\hbox{$\inbar\kern-.3em{\rm G}$}}
 \def\IGa{\relax\hbox{${\rm I}\kern-.18em\Gamma$}}
 \def\IH{\relax{\rm I\kern-.18em H}}
 \def\II{\relax{\rm I\kern-.18em I}}
 \def\IK{\relax{\rm I\kern-.18em K}}
 \def\IP{\relax{\rm I\kern-.18em P}}

\def\Tr{{\rm Tr}}
 \font\cmss=cmss10 \font\cmsss=cmss10 at 7pt
 \def\IR{\relax{\rm I\kern-.18em R}}

\def\a{\alpha}
\def\b{\beta}

\def\D{\Delta}

\def\g{\gamma}

\def\e{\epsilon}

\def\m{\mu}
\def\n{\nu}
\def\s{\sigma}

\def\p{\pi}

\renewcommand{\@}[1]{\sqrt{#1}}
\renewcommand{\le}[1]{\label{#1}\end{eqnarray}}

\def\ffract#1#2{\raise .35 em\hbox{$\scriptstyle#1$}\kern-.25em/
\kern-.2em\lower .22 em \hbox{$\scriptstyle#2$}}

\newdimen\tableauside\tableauside=1.0ex
\newdimen\tableaurule\tableaurule=0.4pt
\newdimen\tableaustep
\def\phantomhrule#1{\hbox{\vbox to0pt{\hrule height\tableaurule width#1\vss}}}
\def\phantomvrule#1{\vbox{\hbox to0pt{\vrule width\tableaurule height#1\hss}}}
\def\sqr{\vbox{%
  \phantomhrule\tableaustep
  \hbox{\phantomvrule\tableaustep\kern\tableaustep\phantomvrule\tableaustep}%
  \hbox{\vbox{\phantomhrule\tableauside}\kern-\tableaurule}}}
\def\squares#1{\hbox{\count0=#1\noindent\loop\sqr
  \advance\count0 by-1 \ifnum\count0>0\repeat}}
\def\tableau#1{\vcenter{\offinterlineskip
  \tableaustep=\tableauside\advance\tableaustep by-\tableaurule
  \kern\normallineskip\hbox
    {\kern\normallineskip\vbox
      {\gettableau#1 0 }%
     \kern\normallineskip\kern\tableaurule}%
  \kern\normallineskip\kern\tableaurule}}
\def\gettableau#1 {\ifnum#1=0\let\next=\null\else
  \squares{#1}\let\next=\gettableau\fi\next}

\newcommand{\auth}{Institute of Theoretical Physics, Aristotle University of Thessaloniki, Thessaloniki, Greece.}
\newcommand{\cern}{Theoretical Physics Department, CERN, Geneva, Switzerland.}

\allowdisplaybreaks
\title{The fermion-boson map for large $d$}
\author[a]{Evangelos G.  Filothodoros,}
\author[b]{Anastasios C. Petkou\footnote{On leave of absence from the Institute of Theoretical Physics, Aristotle University of Thessaloniki, Thessaloniki, Greece.}}
\author[a]{and \,Nicholas D. Vlachos}
\affiliation[a]{\auth}
\affiliation[b]{\cern}

\emailAdd{efilotho@physics.auth.gr}
\emailAdd{tassos.petkou@cern.ch}
\emailAdd{vlachos@physics.auth.gr}
\abstract{We show that the three-dimensional map between fermions and bosons at finite temperature generalises for all odd dimensions $d>3$. We further argue that such a map has a nontrivial large $d$ limit.   
Evidence comes from studying  the gap equations, the free energies and the partition functions  of the $U(N)$ Gross-Neveu and CP$^{N-1}$ models for odd $d\geq 3$ in the presence of imaginary chemical potential. We find that the gap equations and the free energies can be written in terms of the Bloch-Wigner-Ramakrishnan $D_d(z)$ functions analysed by Zagier. Since $D_2(z)$ gives the volume of ideal tetrahedra in 3$d$ hyperbolic space our three-dimensional results are related to resent studies of complex Chern-Simons theories, while for $d>3$ they yield corresponding higher dimensional generalizations.  As a spinoff, we observe that particular complex saddles of the partition functions correspond to the zeros and the extrema of the Clausen functions $Cl_d(\theta)$ with odd and even index $d$ respectively. These saddles lie on the unit circle at positions remarkably well approximated by a sequence of rational multiples of $\p$. }

\begin{document}
\begin{flushright}
CERN-TH-2018-049\\
  \end{flushright}
\maketitle

\section{Introduction}

Three-dimensional bosonization \cite{Fradkin:1994tt} via  statistical transmutation \cite{Wilczek:1981du,Polyakov:1988md} is a recurrent  theme in field theory and condensed matter physics, and it has been more recently connected (see e.g. \cite{Karch:2016sxi,Murugan:2016zal,Seiberg:2016gmd,Kachru:2016rui}) to particle-vortex duality e.g. \cite{Peskin:1977kp}. The extension of fermion-boson duality to finite temperature has also been  considered  (see  for example \cite{Giombi:2011kc,Aharony:2012ns} and references therein), in the context  of various models that describe matter coupled to non-abelian Chern-Simons fields.  

If fermion-boson duality is a fundamental property of three-dimensional quantum physics one may think that it is not necessary to invoke non-abelian gauge fields to make it manifest. Having this idea in mind we revisited in  \cite{Filothodoros:2016txa}  the finite temperature phase structure of two 3$d$ systems; the fermionic $U(N)$ Gross-Neveu model and the bosonic CP$^{N-1}$ model. We have studied those systems in the canonical formalism by introducing an imaginary chemical potential for the $U(1)$ charge. In such a setup the large-$N$ canonical partition functions are intimately related to the  partition functions of the same systems coupled to an abelian Chern-Simons gauge field  expanded around a monopole background in a suitable mean field approximation \cite{Barkeshli:2014ida}. Also, the imaginary $U(1)$ charge density is related to the Chern-Simons level.

It is well known that in the absence of a chemical potential the two models above exhibit very different patterns of symmetry breaking at finite temperature $T$. 
We nevertheless we have shown in \cite{Filothodoros:2016txa} that the situation changes in the presence of the imaginary chemical potential, and the corresponding phase structures of the two models can be mapped into each other. We have further observed in \cite{Filothodoros:2016txa}  the relevance of the celebrated Bloch-Wigner function \cite{Zagier1} in our calculations of the gap equations and free energies. This mathematical curiosity, together with the aim to shed more light into the physics of bosonisation, prompted us to study in the present work the fermion-boson map in higher dimensions.

We begin with a detailed review of the three-dimensional results of \cite{Filothodoros:2016txa}. In particular, we discuss that when we introduce an imaginary chemical potential the phase structures of the Gross-Neveu and CP$^{N-1}$ models are characterised by the presence of {\it thermal windows} inside which the systems do not have definite fermionic or bosonic properties. The edges of the thermal windows are inflection points of the free energies, while in the middle points of the above windows the systems appear to have completely switched statistics i.e. the fermionic becomes bosonic and vice versa. We  show that there exists an explicit map between the two models by simply twisting the boundary conditions in one of the two i.e. in the fermionic one. This way the corresponding gap equations, free energies and consequently the phase structures are mapped into each other. 

The 3$d$ fermion-boson map is translated into the relationship (\ref{pfduality3}) between the corresponding large-$N$ partition functions. And here is where the appearance of the Bloch-Wigner function $D(z)$  gives an important clue. Noticing that $D(z)$ enters in the exponent of the partition function duality relation (\ref{pfduality3}) we begin to suspect that our excursion into the imaginary chemical potential regime of our simple systems may have unveiled some properties of more complicated systems e.g. matter coupled to complex non-abelian Chern-Simons. Indeed, the saddle points at the edges of our thermal windows give {\it exactly} half of the hyperbolic volume of the knot complement that was used in \cite{Gukov:2016njj,Gang:2017hbs} as the "classical" starting point of the perturbative expansion for the partition function of $SL(2,{\mathbb C})$ Chern-Simons theory. 

Encouraged by the above 3$d$ results we move on to study the Gross-Neveu and CP$^{N-1}$ models for odd dimensions $d>3$. Even dimensions are not included in our study since the physics there is much more complicated due to the presence of all kinds of gauge fields. Since the models are nonrenormalizable for $d>3$, we find new power like divergences in the gap equations and the free energies. These cannot be dealt with by simply adjusting the single coupling of the theory, as we did in $d=3$. This  makes ambiguous the detailed analysis of the phase structure of the higher dimensional models. However, we observe that the finite parts of the gap equations and the free energies are given in terms of the generalized Bloch-Wigner-Ramakrishnan function $D_d(z)$ introduced by Zagier \cite{Zagier2}. It is remarkable that the finite parts the divergent integrals conspire to give such a simple result. 

The analysis of the higher dimensional models is more reliable on the unit circle in the complex imaginary potential plane, as there the bosonic or fermionic condensates are zero. In this case we find a phase structure that is similar to the 3$d$ thermal windows, however a more detailed analysis is not possible. Nevertheless, the edges of the higher-dimensional thermal windows are also inflection points of the free energy and they correspond to zeros and extrema of the $D_d(z)$ functions with odd and even $d$ respectively. This pattern continues for all $d$. We are then able to give an analytic formula for the approximate positions of those saddle points in the form of a sequence of rational multiples of $\p$ that lie between $\pi/3$ and $\pi/2$. The relevant numerical results are given in Table 1 and Fig. 1. As $d\rightarrow\infty$ those points accumulate near $\pi/2$ and the relevant $D_d(e^{i\theta})$ function becomes simply $\sin(\theta)$. This observation allows us to ask whether the fermion-boson map has a well defined large $d$ limit, and indeed we argue that for models with supersymmetric matter content the partition function duality formula (\ref{pfdualityd0}) goes to the simpler one (\ref{pfsusylimit}).

In Section 2 we discuss some general aspects of three-dimensional systems at imaginary chemical potential and their relation to abelian Chern-Simons theories coupled to matter and to statistical transmutation. In Section 3 we review in detail the 3$d$ results of \cite{Filothodoros:2016txa} and setup the higher dimensional calculations. In Section 4 we present our results for the gap equations and the free energies of the higher-dimensional models, we discuss the relevance of the $D_d(z)$ functions in our calculations and we give a formula  the approximate positions of the zeros and the extrema of the latter. We summarise and offer a few ideas for future work in Section 5. Three Appendices contain some technical details and useful formulae.

\section{Imaginary chemical potential for global $U(1)$  charges}

\subsection{The canonical ensemble}

Consider a system at finite temperature $T=1/\beta$ with a global $U(1)$ charge operator $\hat{Q}$. Its canonical partition function\footnote{We  normalize $Z_c$ by dividing with the usual thermal partition function $Z(\b)=\Tr e^{-\b\hat{H}}$ such that in the absence of charged states $Z_c(\b,0)=1$.}  is the thermal average over states of fixed $\hat{Q}$ as
\be
\label{canPF}
Z_c(\b,Q)=\Tr\left[\delta(Q-\hat{Q})e^{-\b\hat{H}}\right]\,.
\ee
 Assuming that the eigenvalues $Q$ of the charge operator $\hat{Q}$ are integers, a representation of  (\ref{canPF}) can be written as
\be
\label{gcPF}
Z_c(\b,Q)=\int_{0}^{2\pi}\!\frac{d\theta}{2\pi}\,e^{i\theta Q}\,\Tr\left[e^{-\b\hat{H}-i\theta\hat{Q}}\right]=\int_{0}^{2\pi}\!\frac{d\theta}{2\pi}\,e^{i\theta Q}\,Z_{gc}(\b,\mu=-i\theta/\b)\,,
\ee
where $Z_{gc}(\b,\mu)$ is the grand canonical partition function with imaginary chemical potential $\mu$. The latter function exhibits generically certain periodicity properties wrt $\theta$ that are intimately connected to the physics of the underlying theory. For example \cite{Roberge}, in $SU(N)$ gauge theories with  $\hat{Q}$ the fermion number operator, one expects that in the confining phase  the spectrum contains only colour singlets. In that case $Q$ is a multiple of $N$ and $Z_{gc}(\beta,\mu)$ will be periodic  with a $\theta$-period  $2\pi /N$. If however there is a phase, e.g. at high temperatures, where colourful fundamental particles turn up in the spectrum, one  might expect to find instead a $2\pi$ $\theta$-period.  Indeed, although the $\mathbb{Z}_N$ symmetry of the pure $SU(N)$ Yang-Mills action appears to enforce the $2\pi /N$ periodicity, one generically finds a more complicated structure at high temperature which may be attributed to a deconfining transition (se e.g.  \cite{Aarts:2015tyj}.)

Integrals like (\ref{gcPF}) can be evaluated  by a saddle point analysis. The saddle point equation is
\be
\label{saddlepoint}
iQ-\b\frac{\partial}{\partial\theta}F_{gc}(\b,-i\theta /\b)= 0\,,
\ee
where the grand canonical potential (i.e. the free energy) is $\b F_{gc}(\b,-i\theta/\b)=-\ln Z_{gc}(\b,-i\theta/\b)$. If the grand canonical partition function is an even function of $\m$ and hence of $\theta$, as in most physically relevant situations (i.e. charge conjugation, $CP$ invariance etc), then real solutions for $Q$  require imaginary $\theta$ and one simply gets back to a real chemical potential \cite{Karbstein:2006er}. On the other hand, for the models that we will study below,  complex saddles of (\ref{saddlepoint}) give imaginary $Q$ and yield terms proportional to the volume of hyperbolic 3-manifolds. One then might conceive that complex saddles are related to three dimensional gravity coupled to matter.

Moreover, it is possible that $F_{gc}(\b,-i\theta/\b)$ has one or more extrema for some real $\theta_*\neq 0$. In such cases, the canonical partition function of the system in the absence of charged excitations $Q=0$ is given, to leading order in some approximation scheme such as large-$N$, by the grand canonical partition function of the same system at fixed imaginary chemical potential $\m_*=-i\theta_*/\b$
\be
\label{approx}
Z_c(\b,0)\approx e^{-\b F_{gc}(\b,-i\theta_* /\b)}\,.
\ee
This is not inconsistent with the normalization of $Z_c(\b,0)$ advocated above since  fixing the imaginary chemical potential of a system at finite temperature is tantamount to statistical transmutation \cite{Petkou:1998wd,Christiansen:1999uv}. Namely, the system described by (\ref{approx}) would be in general different from the initial one as its elementary degrees of freedom would obey different statistics.

\subsection{Matter coupled to Chern-Simons in a monopole background}

It is known that for scalars and fermions are coupled to a gauge field $A_\m$ at finite temperature, the temporal component $A_0$ plays the role of an imaginary chemical potential (see e.g.\cite{Kapusta:2006pm,ZinnJustin:2002ru}). 
In \cite{Filothodoros:2016txa} we have argued that for Dirac fermions coupled to an abelian Chern-Simons field at level $k$ in three Euclidean  dimensions\footnote{Our notations follow \cite{ZinnJustin:2002ru} and are briefly presented in the Appendix.} the presence of an imaginary chemical potential is intimatelly connected to the presence of monopole charge. To see the connection consider the following finite temperature fermionic partition function 
\begin{align}
\label{DiracCSPF}
Z_{f}(\b,k)&=\int [{\cal D}A_\m][{\cal D}\bar\psi][{\cal D}\psi]\exp{\left[-S_f(\bar\psi,\psi,A_\m)\right]}\,,\\
\label{Sf}
S_f(\bar\psi,\psi,A_\m)&=-\int_0^{\beta}\!\!\!dx^0\!\!\int \!\!d^2\bar{x}\left[\bar{\psi}(\slash\!\!\!\partial -i\slash\!\!\!\!A)\psi+i\frac{k}{4\pi}\epsilon_{\m\n\rho}A_\m\partial_\n A_\rho+...\right]\,,
\end{align}
where the dots denote the possible presence of fermionic self interactions. We expand the CS field around a static (i.e. $x^0$-independent) monopole configuration $\bar{A}_\m$ \cite{Fosco:1998cq}
\be
\label{Aexpansion}
A_\m=\bar{A}_\m+\alpha_\m\,,\,\,\,\bar{A}_\m=(0,\bar{A}_1(\bar{x}),\bar{A}_2(\bar{x}))\,,\,\,\, \a_\m=(\a_0(x^0),\a_1(x^0,\bar{x}),\a_2(x^0,\bar{x}))\,,
\ee
normalized as\footnote{For example, one may consider the theory on $S^1\times S^2$.}
\be
\label{monopole}
\frac{1}{2\pi}\int d^2x \bar{F}_{12}=1\,,\,\,\,\bar{F}_{\m\n}=\partial_\m \bar{A}_\n-\partial_\n \bar{A}_\n\,.
\ee
Hence, (\ref{Sf}) describes the attachment of $k$ units of monopole charge to the fermions as
\be
\label{Sfexp}
S_{f}(\bar\psi,\psi,A_\m) =-\int_0^\b \!\!\!dx^0\!\!\int \!\!d^2\bar{x}\left[\bar\psi(\slash\!\!\!\partial-i\gamma_i\bar{A}_i-i\gamma_\m\a_\m)\psi+i\frac{k}{4\pi}\epsilon_{\m\n\rho}\a_\m\partial_\n\a_{\rho}+..\right]-ik\int_0^\b \!\!\!dx^0 a_0\,.
\ee
We can perform the path integral over the CS fluctuations projecting into a sector with fixed total monopole charge. To do so, we assume the existence of a mean field approximation within this sector such that the spatial CS fluctuations compensate for the magnetic background  $\langle \a_i\rangle =-\bar{A}_i$ \cite{Barkeshli:2014ida}.\footnote{The validity of such an approximation probably requires also a suitable large-$N$ limit and it would be interesting to clarify it further.} We then obtain
\begin{align}
Z_{f}(\b,k)&=\int [{\cal D}\a_0][{\cal D}\bar\psi][{\cal D}\psi]\exp{\left[\int_0^\b \!\!\!dx^0 \!\!\int \!\!d^2\bar{x}\left[\bar\psi(\slash\!\!\!\partial-i\gamma_0\a_0)\psi+..\right]+ik\int_0^\b \!\!\!dx^0\a_0\right]}\nonumber \\
\label{DiracCSPFfin}
&=\int ({\cal D}\theta)e^{ik\theta}Z_{gc,f}(\b,-i\theta/\b)\,,
\end{align}
where  $\theta=\int_0^\b dx^0\a_0(x^0)$ and we have used standard formulae from \cite{Kapusta:2006pm,ZinnJustin:2002ru}. Comparing  with (\ref{saddlepoint}) we see that the CS level $k$ plays the role of the eigenvalue $Q$ of the $U(1)$ charge operator. 

Similarly, the thermal partition function of a complex scalar $\phi$ coupled to abelian CS at level $k$ may be written as
\begin{align}
\label{ScalarCSPF}
Z_{b}(\b,k)&=\int [{\cal D}A_\m][{\cal D}\bar\phi][{\cal D}\phi]\exp{\left[-S_{b}(\bar\phi,\phi,A_\m)\right]}\,,\\
\label{Ssc}
S_{b}(\bar\phi,\phi,A_\m) &=\int_0^\b\!\! \!dx^0 \!\!\int \!\!d^2\bar{x} \left[|(\partial_\m-iA_\m)\phi|^2 -i\frac{k}{4\pi}\epsilon_{\m\n\rho}A_\m\partial_\n A_\rho+..\right]\,,
\end{align}
where again with the dots we have allowed for the presence of a non trivial scalar potential. Expanding as in (\ref{Aexpansion}), (\ref{monopole}) and assuming a similar mean field  and large-$N$ approximation  we find
\begin{align}
\label{ScalarCSPFfin}
Z_{b}(\b,k)&=\int[{\cal D}\a_0][{\cal D}\bar{\phi}][{\cal D}\phi]\exp{\left[-\int_0^\b \!\!\!dx^0 \!\!\int \!\!d^2\bar{x}\left[|(\partial_x^0-i\a_0)\phi|^2 +|\partial_i\phi|^2 +..\right]+ik\int_0^\b \!\!dx^0 a_0\right]} \nonumber \\
&=\int [{\cal D}\theta]e^{ik\theta}Z_{gc,b}(\b,-i\theta/\b)\,,
\end{align}
where we have used the same definition for $\theta$ as above, and compared with the standard formulae giving the grand canonical partition function for charged scalars \cite{Kapusta:2006pm,ZinnJustin:2002ru}. 

\subsection{Statistical transmutation}

Finally we briefly review the relationship of the imaginary chemical potential to statistical transmutation. Consider for example the fermionic theory (\ref{DiracCSPFfin}). One notices that the presence of the imaginary chemical potential can be cancelled by the following abelian gauge transformation for the fermions
\be
\label{fgaugetransf}
\psi(x^0,\bar{x})\mapsto \psi'(x^0,\bar{x})=e^{i\int_0^{x^0} d\tilde{x}^0\alpha_0(\tilde{x}^0)}\psi(x^0,\bar{x})\,,\,\,\,\bar\psi(x^0,\bar{x})\mapsto \bar\psi'(x^0,\bar{x})=e^{-i\int_0^{x^0} d\tilde{x}^0\alpha_0(\tilde{x}^0)}\bar\psi(x^0,\bar{x})\,.
\ee
However, at finite temperature the fermions obey anti-periodic boundary conditions on the thermal circle
\be
\label{apBC}
\psi(\beta,\bar{x})=-\psi(0,\bar{x})\,,\,\,\,\bar{\psi}(\beta,\bar{x})=-\bar{\psi}(0,\bar{x})\,.
\ee
We then see that the gauge transformed fields would satisfy
\be
\label{apBC1}
\psi'(\beta,\bar{x})=-e^{i\theta}\psi'(0,\bar{x})\,,\,\,\,\,\bar\psi'(\b,\bar{x})=-e^{-i\theta}\bar\psi'(0,\bar{x})\,.
\ee
Hence, the anti-periodic boundary conditions are preserved only if $\theta=2\pi n$, $n\in \mathbb{Z}$. Other values of $\theta$ would "twist" the boundary conditions and change the  statistical properties of the undelying system. A similar argument goes  through  for  bosonic systems where the complex scalars satisfy periodic boundary conditions on the thermal circle. The twisting of the thermal boundary conditions is the main underlying mechanics behind the possible statistical transmutation in systems whose grand canonical potential is extremized at non trivial values of the imaginary chemical potential.

\section{Revisiting the $3d$ fermion-boson map at imaginary chemical potential}

The calculation of the canonical partition function (\ref{gcPF}) in systems with  global $U(1)$ charges appears to be agnostic to their underlying microscopic structure e.g. whether the elementary degrees of freedom carrying charge are bosonic or fermionic.  Indeed, it looks like that the only useful piece of information one has is the kind of periodicity of the  partition function, something that could just give a hint regarding the presence of a confinement/deconfinement transition. 

The situation resembles studies of quantum mechanical systems in a periodic potentials. If we think of $\theta$ as a periodic coordinate, then (\ref{gcPF}) is equivalent to the calculation of the overlap between two Bloch wavefunctions that differ by  lattice momentum $Q$ (see e.g. eq. 2 of \cite{Blochoverlaps}). Although one generally cannot go very far without using a particular microscopic model at hand, there are certain topological properties of a single band such the eigenvalues of the Zak phase \cite{Zak}, which hold physically relevant information of the system i.e. polarization.  We will not pursue  further this line of ideas here, but if there is a lesson to be learned is that there may be some universal features of generic quantum systems in periodic potentials which are independent of their fermionic or bosonic microstructure. Hence, a fermion-boson map appears to be generic.  This will be exploited below by considering two explicit three dimensional models: the $U(N)$ fermionic Gross-Neveu and the bosonic CP$^{N-1}$  model.

\subsection{The $U(N)$ Gross-Neveu model at imaginary chemical potential for $d=3$}

To calculate the canonical partition function of the $U(N)$ Gross-Neveu model in the presence of imaginary chemical potential $\m=-i\alpha$ we use the Euclidean action \cite{Petkou:1998wd,Christiansen:1999uv}
\begin{equation}
\label{GNaction}
S_{GN} = -\int_0^\b \!\!\!dx^0\int \!\!d^2\bar{x} \left[\bar{\psi }^{a}(\slash\!\!\!\partial  -i\gamma_0\alpha)\psi ^{a}+\frac{G_3}{2({\rm Tr}\mathbb I_2)N}\left (\bar{\psi }^{a}\psi ^{a}\right )^{2} +i\a NQ_3\right]\,,\,\,\,a=1,2,..N\,.
\end{equation}
where $Q_3$ is the eigenvalue of the $N$-normalized fermion number density\footnote{Wrt  the spatial volume $V_2$.} operator $\hat{Q}_3=\psi^{a\dagger}\psi^a/N$ in $d=3$. Introducing an auxiliary scalar field $\sigma$ the canonical partition function is given by
\begin{align}
\label{GNPF1}
Z_f(\b,Q_3)&=\int({\cal D}\a)({\cal D}\sigma)e^{-N S_{f,eff}}\,,\\
\label{GNPF2}
S_{f,eff}&=iQ_3\int_0^\b \!\!\!dx^0\!\!\int\!\!d^2\bar{x}\,\alpha -\frac{{\rm Tr}\mathbb I_2}{2G_3}\int_0^\b \!\!\!dx^0\!\!\int d^2\bar{x}\,\sigma^2+\Tr\ln\left(\slash\!\!\!\partial-i\gamma_0\alpha+\sigma\right)_\b\,.
\end{align}
\subsubsection{The gap equations}
To evaluate (\ref{GNPF1}) we look for constant saddle points $\alpha_*$ and $\sigma_*$. At large-$N$ these are given by the gap equations
\begin{align}
\label{GNgap1}
\frac{\partial}{\partial\sigma}S_{f,eff}\Biggl|_{\sigma_*,\a_*}\!\!\!\!=0\,\,\,\Rightarrow\,\,\, -\frac{\sigma_*}{G_3}&+\frac{\sigma_*}{\b}\sum_{n=-\infty}^\infty\int^\Lambda\!\!\frac{d^2 \bar{p}}{(2\pi)^2}\frac{1}{\bar{p}^2+(\omega_n-\alpha_*)^2+\sigma_*^2}=0\,,\\
\label{GNgap2}
\frac{\partial}{\partial\a}S_{f,eff}\Biggl|_{\sigma_*,\a_*}\!\!\!\!=0\,\,\,\Rightarrow\,\,\, i Q_3&-\lim_{\epsilon\rightarrow 0}\frac{{\rm Tr}\mathbb I_2}{\b}\int^\Lambda\!\!\frac{d^2 \bar{p}}{(2\pi)^2}\sum_{n=-\infty}^\infty\frac{e^{i\omega_n\epsilon}(\omega_n-\alpha_*)}{\bar{p}^2+(\omega_n-\alpha_*)^2+\sigma_*^2}=0\,,
\end{align}
where  the fermionic Matsubara sums are over the discrete frequencies $\omega_n=(2n+1)\pi/\b$. The divergent integrals are regulated by the cutoff $\Lambda$. As explained in Appendix A, we have used the parameter $\epsilon$ to regulate the sum in (\ref{GNgap2}) before performing the integral. 


In \cite{Filothodoros:2016txa} we noticed that the above gap equations are nicely expressed in terms of the Bloch-Wigner-Ramakrishna $D_d(z)$ functions\footnote{We collect some properties of the $D_d$ function in Appendix B.}  analysed by Zagier in \cite{Zagier1} as
\begin{align}
\label{gap31}
&\sigma_*\left[-{\cal M}_3\beta+D_1(-z_*)\right]=0\,,\\
\label{gap32}
&\frac{2\pi}{{\rm Tr}\mathbb I_2}\beta^2Q_3+iD_2(-z_*)\,,=0
\end{align}
where $z_*=e^{-\beta\sigma_*-i\beta\alpha_*}$. The $D_d(z)$'s are real valued complex functions, hence  $Q_3$ is purely imaginary. Moreover, the Bloch-Wigner function $D(z)\equiv D_2(z)$ gives the volume of ideal tetrahedra in Euclidean hyperbolic space ${\cal H}_3$ whose four vertices lie in $\partial{\cal H}_3$ at the points $ 0, 1,\infty,$ and $z$ ($z$ is a dimensionless cross ratio here). These tetrahedra are the building blocks for general hyperbolic manifolds - the volume of the latter arises as the sum of ideal tetrahedra after a suitable triangulation \cite{Zagier1}. We now note a remarkable connection of our fermionic model with imaginary chemical potential to three-dimensional gravity.  Namely, it is known \cite{Witten:1989ip,Gukov:2003na} that complex $SL(2,\mathbb{C})$ Chern-Simons theory with purely imaginary level corresponds, at least semi classically,  to Euclidean three-dimensional gravity with negative cosmological constant. As $Q_3$ corresponds to a purely imaginary Chern-Simons level according to the discussion of Section 2.2, we can understand the presence of hyperbolic volumes in the gap equations and eventually in the free energy of our model.

The mass scale ${\cal M}_3$ quantifies the distance between the bare the critical couplings as
\be
\label{MGcrit}
\frac{{\cal M}_3}{2\pi}\equiv\frac{1}{G_{3,*}}-\frac{1}{G_{3}},\,\,\,\,\,\,\,\frac{1}{G_{3,*}}\equiv\int^\Lambda\!\!\!\frac{d^3p}{(2\pi)^3}\frac{1}{p^2}\,.
\ee
To obtain (\ref{gap31}) and (\ref{gap32}) we have dropped terms that vanish as the cutoff is sent to infinity. It should also be noted that the cut-off dependance drops out in the second of the gap equations (\ref{gap31}). The explicit calculations leading to the gap equations above, for general $d$, are presented in the Appendix A.

\subsubsection{The phase structure}
The physics of the model is read from the above gap equations.  A non-zero solution for $\sigma_*$  implies parity symmetry breaking as it corresponds to a nonzero mass for the elementary fermions. 
\begin{itemize}
\item \underline{$\a_*=0$}
\end{itemize}
Since $D_1(-z)<0$ for $z\in\mathbb{R}$, parity symmetry breaking is only possible  when ${\cal M}_3>0$. This requires strong coupling $G_{3}>G_{3,*}$. Nevertheless 
we can always find a high enough temperature $T=1/\beta$ satisfying $\beta{\cal M}_3\leq \ln 2$ at which parity will be restored again since $\sigma_*$ will vanish. For a given ${\cal M}_3\neq 0$ the critical temperature for parity restoration is $T_c={\cal M}_3/\ln 2$. In the weak coupling regime when $G_3\leq G_{3,*}$ parity is never broken and the only solution of (\ref{gap31}) is $\sigma_*=0$. 

An equivalent description of the physics of parity breaking is as follows. Starting at some high enough temperature $T=1/\beta$ where parity is intact we can ask how deep we should need to go into the strong coupling regime to break it. The answer is  that we need to tune $G_3$ such that ${\cal M}_{3}\geq T\ln 2$. The equality actually gives the (subtracted - see  Appendix C) free energy density for massless free fermions in $d=1$. In other words, finite temperature introduces a barrier equal to the energy needed to excite one {\it $1d$} massless fermion for each 3d fermionic d.o.f.\footnote{This follows since our quantities are $N$-normalised.} which must be overtaken if we want to break parity in  three dimensions. 

Moreover, using 
\be
\label{dD1}
\frac{\partial}{\partial z}D_1(z)=-\frac{1+z}{4z(1-z)}\,,
\ee
we can take the second derivative\footnote{Note the useful formulae $\partial/\partial\s=-\b(z\partial_z+\bar{z}\partial_{\bar{z}})$ and $\partial/\partial\a=-i\b(z\partial_z-\bar{z}\partial_{\bar{z}})$.} of the effective action, which is equivalent to taking the first derivative of the gap equation (\ref{gap31}), and obtain up to an overall constant
\be
\label{ddSeff}
\frac{\partial^2}{\partial\sigma^2}S_{f,eff}\Biggl|_{\sigma_*,\a_*}\propto-{\cal M}_3\b+D_1(-z_*)+\frac{\ln|z_*|}{2}\Re\left(\frac{z_*-1}{z_*+1}\right)\,.
\ee
For $\a_*=0$ the concavity of the effective action at the point $\s_*=0$ depends on whether the coupling ${\cal M}_3$ is bigger or smaller than $TD_1(-1)=T\log 2$. Hence, tuning the coupling to ${\cal M}_3=T\ln 2$ corresponds to the condition that $\s_*=0$ becomes an inflection point of the effective action. At this point, its leading behaviour is as $S_{f,eff}\sim \b^2 V_2\sigma^4$, since from (\ref{Dd1}) we see that $D_1(-z)$ is an even function of $\b\s$. It should be emphasized, however, that although the presence of an infection point for the effective action is an indication for the existence of a phase transition or a crossover, this need not be necessarily the case. In the three-dimensional model above we are able to confirm the finite temperature parity breaking/restoring transition via a detailed analysis of the gap equations. We will see that this is not obvious for $d>3$, nevertheless the presence of inflection points in the effective action will be one of the few robust results that we are able to obtain for the higher-dimensional models.


\begin{itemize}
\item \underline{$\a_*=-i\mu\neq 0$ with $\mu\in \mathbb{R}$}
\end{itemize}
Real chemical potential $\m$ was studied in various works in the past e.g. \cite{Hands:1992ck}. In this case parity is  restored more easily starting inside the strong coupling regime ${\cal M}_3>0$. In fact, there exists a critical value for $\m$ such that parity symmetry is already restored at zero temerpature. Beyond this critical chemical potential parity is never broken  \cite{Christiansen:1999uv}. Moreover, since $D_2(z)$ vanishes for real $z$, the purely imaginary charge $Q_3$ is zero.

A salient property of the fermionic GN model is the absence of a parity symmetry breaking phase at the critical point  ${\cal M}_3=0$, with or without a {\it real} chemical potential. This changes dramatically in the presence of imaginary chemical potential. The reason is that the function $D_1(-z)$ has nontrivial zeros when $2\pi/3\leq {\rm Arg} z\leq 4\pi/3$   \footnote{Since we take $\a_*>0$, we consider $\b\a_*\in [0,2\pi]$ to ensure a positive temperature $T$.} and hence the gap equation (\ref{gap31})  can be satisfied for $\s_*\neq 0$. The zeros of $D_1(-z)$ are given by the  roots of the quadratic equation  \cite{Christiansen:1999uv}
\be
\label{roots3df}
x^2+(2\cos\b\alpha_*-1)x+1=0\,,\,\,\,x=e^{-\b\sigma_*}\,.
\ee
The roots are real  when $1/A_*< 1/T< 2/A_*$ (mod $2\p/\a_*$), where $A_*=3\a_*/2\p$. In particular,
\begin{itemize}

\item \underline{$\a_*\neq 0$ and $A_*\leq T <\infty$,\,\, $\frac{A_*}{3k+1}\leq T\leq \frac{A_*}{3k-1}$, with $k=1,2,..$}
\end{itemize}
Consider the critical system with ${\cal M}_3=0$. At high enough temperature parity is always unbroken. We may try to break it by lowering $T$ but this can never happen inside the temperature windows  above. There, the only solution to the gap equations is $\sigma_*=0$. 
\begin{itemize}
\item \underline{$\a_*\neq 0$ and $\frac{A_*}{3k+2}\leq T\leq \frac{A_*}{3k+1}$, with $k=0,1,2,..$}
\end{itemize}
For the complementary temperature ranges above - which we term {\it bosonic thermal windows of the GN model} - the critical gap equation {\sl does} have solutions with $\sigma_*\neq 0$. This property is alien to a {\it critical fermionic} model, and rather resembles a property of a {\it critical  bosonic} model at finite temperature \cite{Sachdev:1993pr,Petkou:1998wd} as we will review below. 
\begin{itemize}
\item \underline{The edges of the thermal windows}
\end{itemize}
The edges of the above  thermal windows correspond to the roots of $D_1(-z)$ on the unit circle
\be
\label{D1roots}
D_1(-e^{-i\beta\alpha_*})\equiv Cl_1(\b\a_*\pm\p)=0\,\,\,\Rightarrow\,\,\,\beta\alpha_*=\frac{2\pi}{3}\,\, {\rm or}\,\, \b\a_*=\frac{4\pi}{3}\,\, ({\rm mod}\,2\pi)\,.
\ee
Notice the property of the Clausen functions $Cl_{m}(\theta)$ with odd indexes
\be
\label{ClausenOdd}
Cl_{2n-1}(\theta\pm\pi)=-Cl_{2n-1}(\theta)\,,\,\,\,n=1,2,3..\,.
\ee
Since for $|z|=1$ it holds  \cite{Zagier1}
\be
\label{DmDm-1}
\frac{\partial}{\partial z}D_d(z)=\frac{i}{2z}D_{d-1}(z)\,,\,\,\,d=2,3,...\,,
\ee
the edges of the bosonic thermal windows of the GN model are the extrema of $D_2(-z)$ and hence of the purely imaginary charge $Q_3$.  Indeed, using the periodicity properties of the Clausen function with even index
\be
\label{ClausenEven}
Cl_{2n}(2\pi-\theta)=-Cl_{2n}(\theta)\,,\,\,\,n=1,2,3,..\,,
\ee
we find
\be
 \label{Qmax}
 \frac{2}{{\rm Tr}\mathbb I_2}Q_{3,extr}=\mp\frac{i}{\pi\b^2}Cl_2\left(\frac{\pi}{3}\right)\,.
 \ee
 for $\b\a_*=2\pi/3$ and $4\pi/3$ respectively. $Cl_2(\pm \pi/3)=\pm1.01494$  is the global maximum (minimum) of the Clausen function.  

Finally, by virtue of  (\ref{ddSeff})  we see that the edges of the thermal windows are also inflection points of the critical (${\cal M}_3=0$) effective action.  This fits with our result that connects those points with a crossover between a fermionic and a bosonic behaviour. We will find below similar inflection points on the of unit circle of the critical effective action in higher dimensions, however an analytic proof that they correspond to some kind of crossover behaviour is not yet there.


\begin{itemize}
\item \underline{$Q_3=0$ for $\a_*\neq 0$}
\end{itemize}
At the middle point of the thermal windows $\b\a_*=\pi$ (mod $2\p$) we have $Q_3=0$ and hence there are no charged excitations. The system now has been completely bosonised, and for instance the {\it critical}  gap equation (\ref{gap31}) becomes
\be
\label{D1}
\frac{\s_*}{\p\b}D_1(e^{-\b\s_*})=0\,,
\ee
This is exactly the {\it critical} bosonic gap equation, to be discussed below, multiplied by $\sigma_*$. But now, the gap equation has two real solutions: the trivial one $\sigma_*=0$ and the nonzero root of $D_1(\Re(z))$ which is given by 
\be
\label{gmean}
\beta\sigma_*=2\ln\left(\frac{1+\sqrt{5}}{2}\right)\approx 0.962424.. \,.
\ee

\subsubsection{The  free energy density}

In \cite{Filothodoros:2016txa} we have calculated the large-$N$ subtracted  free energy density of the model at $G_3=G_{3,*}$ as
\be
\label{freeEf}
\frac{2}{N(\Tr{\mathbb{I}_2})}\D f^{(3)}_f(\b)\equiv \frac{2}{N(\Tr{\mathbb{I}_2})}\Bigl(f_f(\infty) -f_f(\b)\Bigl)=-\frac{\sigma_*^2}{G_{3,*}}+i\frac{2}{\Tr{\mathbb{I}_2}}\a_*Q_3-\frac{1}{\p\b^3}D_3(-z_*)\,,
\ee
where  $f(\b)=-\ln Z(\b)/(\b V_2)$ and we have used the charge gap equation (\ref{gap32}). This is a real quantity. To further make sense of it we need to subtract the linear divergence due to $1/G_{3,*}$. Doing so, in the physically clearer case when $Q_3=0$ we distinguish three cases. When $\a_*=0$ the gap equation gives $\sigma_*=0$ and hence, for both the free  ($G_3=0$) and the critical ($G_3=G_{3,*}$) theories we find
\be
\label{freeEf0}
\frac{2}{N(\Tr{\mathbb{I}_2})}\Delta f^{(3)}_f(\b)\Bigl|_{\b\a_*=0} = \frac{2}{N(\Tr{\mathbb{I}_2})}\Delta f^{(3)}_{f,free}(\b) = \frac{3}{4}\frac{\zeta(3)}{\pi\beta^3}\,,
\ee
which coincides with the free energy density of a massless Dirac fermion in $d=3$. When $\b\a_*=\p$ we have from the one hand the solution $\sigma_*=0$ and on the other the logarithm of the golden mean (\ref{gmean}), which yield 
\be
\label{freeEf1}
\frac{2}{N(\Tr{\mathbb{I}_2})}\Delta f^{(3)}_f(\b)\Bigl|_{\b\a_*=\p} = \left\{\begin{array}{ll}-\frac{\zeta(3)}{\pi\beta^3}\,,&\s_*=0\\ -\frac{4}{5}\frac{\zeta(3)}{\pi \b^3}\,,&\beta\sigma_*=2\ln\left(\frac{1+\sqrt{5}}{2}\right)\end{array}\right.\,.
\ee
The above two  cases are {\it minus} the corresponding free energy densities of the bosonic model discussed below at its free and interacting critical points \cite{Sachdev:1993pr,Petkou:1998wd}.  At the edges of the thermal windows we find
\begin{align}
\label{GNfeWindows1}
\frac{2}{N(\Tr{\mathbb{I}_2})}\Delta f_{f}^{(3)}(\b)\Bigl|_{\b\a_*=2\p/3}&=-\frac{1}{3\p\b^3}\left[\zeta(3)-2\pi Cl_2\left(\frac{\p}{3}\right)\right]\,,\\
\label{GNfeWindows2}
\frac{2}{N(\Tr{\mathbb{I}_2})}\Delta f_{f}^{(3)}(\b)\Bigl|_{\b\a_*=4\p/3}&=-\frac{1}{3\p\b^3}\left[\zeta(3)+ 4\pi Cl_2\left(\frac{\p}{3}\right)\right]\,.
\end{align}
It is remarkable that for $\b\a_*=2\pi/3,4\pi/4$ the function $D_3(-e^{-i\b\a_*})$ is a rational multiple of $\zeta(3)$, explicitly $D_3(-e^{-i2\pi/3})=\zeta(3)/3$. This results first appeared in \cite{Christiansen:1999uv} and also later in \cite{LeClair:2004bd} where it was used to classify possible rational CFTs in dimensions $d>2$.  The results (\ref{GNfeWindows1}) and  (\ref{GNfeWindows2}) resemble the sum of contributions coming from conformal matter  (the $\zeta(3)$ term), and a hyperbolic volume  (the $Cl_2(\p/3)$ term).

\subsection{The CP$^{N-1}$ model at imaginary chemical potential for $d=3$}

The Euclidean action for the CP$^{N-1}$ model \cite{Arefeva:1980ms,DiVecchia} with an imaginary chemical potential is
\be
\label{CPaction}
S_{CPN}=\int_0^\b \!\!\!dx^0 \!\!\int \!\!d^2\bar{x}\left[|(\partial_0-i\a)\phi^a|^2 +|\partial_i\phi|^2 +i\lambda(\bar{\phi}^a\phi^a-\frac{N}{g_3})+iNq_3\alpha\right]\,,\,\,\,a=1,2,..,N\,,
\ee
where the auxiliary scalar field $\lambda$ enforces the constraint $|\phi|^2=N/g_3$ and $q_3$ is the  eigenvalue density of the $N$-normalized $U(1)$ charge density operator $\hat{q}_3=-ig\bar\phi^a\overset{\leftrightarrow}\partial_0\phi^a/N$. The model has a global $SU(N)$ symmetry, as well as a global $U(1)$ symmetry that can be trivially gauged by the introduction of a non-propagating abelian gauge field. Integrating out the scalar fields we obtain the canonical partition function as
\begin{align}
\label{CPPF1}
Z_b(\b,q)&=\int({\cal D}\alpha)({\cal D}\lambda)e^{-NS_{b,eff}}\,,\\
\label{CPPF2}
S_{b,eff}&= iq_3\int_0^\b\!\!\!dx^0 \!\!\int \!\!d^2\bar{x}\,\alpha+i\frac{1}{g_3}\int_0^\b\!\!\!dx^0\!\!\int \!\!d^2\bar{x}\,\lambda-\Tr\ln\left(-(\partial_0-i\a)^2-\partial^2+i\lambda\right)_\b\,.
\end{align}
\subsubsection{The gap equations}
Again, we  evaluate (\ref{CPPF1}) at constant saddle points to obtain the gap equations
\begin{align}
\label{CPgap1}
\frac{\partial}{\partial (i\lambda)}S_{b,eff}\Biggl|_{\lambda_*,\a_*}\!\!\!\!=0\,\,\,\Rightarrow\,\,\, \frac{1}{g_3}&-\frac{1}{\b}\sum_{n=-\infty}^\infty\int\frac{d^2 \bar{p}}{(2\pi)^2}\frac{1}{\bar{p}^2+(\omega_n-\alpha_*)^2+m_*^2}=0\,,\\
\label{CPgap2}
\frac{\partial}{\partial\a}S_{b,eff}\Biggl|_{\lambda_*,\a_*}\!\!\!\!=0\,\,\,\Rightarrow\,\,\, iq_3&+\lim_{\epsilon\rightarrow 0}\frac{2}{\b}\int\frac{d^2 \bar{p}}{(2\pi)^2}\sum_{-\infty}^\infty\frac{e^{i\omega_n\epsilon}(\omega_n-\alpha_*)}{\bar{p}^2+(\omega_n-\alpha_*)^2+m_*^2}=0\,,
\end{align}
where the bosonic  frequencies are $\omega_n=2\pi n/\b$ and we have set $i\lambda_*\equiv m_*^2$ in order to facilitate the comparison compare with the fermionic gap equations (\ref{gap31}) and (\ref{gap32}) as $\sigma_*$ and $m_*$ have the same dimensions. 

The gap equations found in \cite{Filothodoros:2016txa} can be written as
\begin{align}
\label{gapb31}
&-{\cal N}_3\beta+D_1(\hat{z}_*)=0\,,\\
\label{gapb32}
&\pi\beta^2q_3-iD_2(\hat{z}_*)=0\,,
\end{align}
where here $\hat{z}_*=e^{-\b m_*-i\b\a_*}$ and the mass scale ${\cal N}_3$ parametrising the difference of $g_3$ from the critical coupling $g_{3,*}$ is defined as
\be
\label{Ncrit}
\frac{{\cal N}_3}{2\pi}\equiv \frac{1}{g_{3,*}}-\frac{1}{g_{3}}\,,\,\,\,\,\,\,\,\frac{1}{g_{3,*}}\equiv\int^\Lambda\!\!\!\frac{d^3p}{(2\pi)^3}\frac{1}{p^2}\,.
\ee
\subsubsection{The phase structure}
\begin{itemize}
\item \underline{$\a_*=0$}
\end{itemize}
The phase structure of the model has been studied in the past at zero temperature e.g. \cite{DiVecchia,Murthy:1989ps}. At finite temperature $D_1(e^{-\b m_*})$ diverges for $m_*=0$ and for arbitrary ${\cal N}_3\neq 0$ there exist a unique nonzero solution to the gap equation (\ref{gapb31}). This yields the bosonic thermal mass which essentially measures how much differs $g_3$ from the critical coupling $g_{3,*}$. It is not possible to find a finite temperature $T>0$ for which the thermal mass vanishes, as that would imply the breaking of the global $O(N)$ symmetry by the Goldstone mechanism \cite{Rosenstein:1989sg}. This is of course consistent with the Mermin-Wagner-Coleman theorem that forbids the breaking of a continuous symmetry in two spatial dimensions.  At the critical point  $g_3=g_{3,*}$ the nontrivial solution to (\ref{gapb31}) is given by the nontrivial root of $D_1(e^{-\b m_*})$ i.e. by logarithm of the  "golden mean" (\ref{gmean}). The existence of this nontrivial critical point at finite temperature is a salient feature of the bosonic model, and should be contrasted with the fermionic case where there was no nontrivial fermion mass at criticality. 
%
\begin{itemize}
\item \underline{$\a_*=-i\mu\neq 0$ with $\mu\in \mathbb{R}$}
\end{itemize}
The presence of a real chemical potential simply shifts the value of the thermal mass. The basic features  of the phase structure of the model remain unaltered.

The question is then whether the phase structure of the model changes in the presence of an imaginary chemical potential. This is equivalent to asking whether $D_1(z_*)$ has roots in the complex plane. As we have discussed above, and we have shown in \cite{Filothodoros:2016txa}, this boils down to solving the quadratic equation 
\be
\label{eq1}
x^2-(2\cos\b\alpha_* -1)x+1=0\,,\,\,\,x=e^{-\b m_*}\,.
\ee
This has real roots for $x$  when  $0<\b <1/2A_*$ and $5/2A_*<\b <3/A_*$ (mod $2\pi/\a*$).  
\begin{itemize}
\item \underline{$\a_*\neq 0$ and $2A_*\leq T<\infty$, \,\,$\frac{2A_*}{5+6k}\leq T\leq \frac{2A_*}{1+6k}$\,,\,\,\,$k=0,1,2,..$}
\end{itemize}
There is always a real nontrivial solution to the {\it critical} gap equation (\ref{gapb31}) when the temperature is tuned inside the intervals given above. Hence, the phase structure retains its bosonic nature.
\begin{itemize}
\item \underline{$\a_*\neq 0$ and $\frac{2A_*}{7+6k}\leq T\leq \frac{2A_*}{5+6k}$\,,\,\,\,$k=0,1,2,..$}
\end{itemize}
The novel feature of the imaginary chemical potential is that there exists an infinite set of {\it fermionic thermal windows of the $CP^{N-1}$ model} given above. Inside these windows there is no nonzero solution to the {\it critical} gap equation (\ref{gapb31}) and hence the critical model cannot acquire a nontrivial thermal mass. Since $D_1(z)\geq 0$ for $\p/3\leq {\rm Arg} z\leq 5\pi/3$, a solution for  (\ref{gapb31}) inside the thermal window requires ${\cal N}_3<0$. This in turn implies that one needs to move into the strong coupling regime $g_3>g_{3,*}$ of the model. Actually, for a given temperature $T$ a nonzero thermal mass can always be found if ${\cal N}_3\leq -T\ln 2$. Hence, in the above thermal window it is always possible to tune the coupling to the special value ${\cal N}_3=-T\ln 2$, such that the thermal mass vanishes. This is reminiscent of the behaviour of the fermionic model where for a given nonzero temperature it was always possible to tune the couplings such as to restore parity and have $\sigma_*=0$. We are therefore tempted to associate the above thermal windows with a fermionic-like phase of the $CP^{N-1}$ model. 

\begin{itemize}
\item \underline{The edges of the thermal windows}
\end{itemize}
The edges of the bosonic thermal windows correspond to the onset of "fermionization". These are given by roots on the unit circle of $D_1(z_*)$ and we find 
\be
\label{D11roots}
D_1(e^{-i\beta\alpha_*})=Cl_1(\b\a_*)=0\,\,\,\Rightarrow\,\,\,\beta\alpha_*=\frac{\pi}{3}\,{\rm or}\, \b\a_*=\frac{5\pi}{3}\,,\,\,\,({\rm mod} \,2\pi)\,.
\ee
Again, the charge is extremized at the edges of the bosonic thermal windows and is given by
\be
\label{qmax}
q_{3,extr}=\mp\frac{i}{\pi\b^2}Cl_2\left(\frac{\pi}{3}\right)\,.
 \ee 
for $\b\a_*=\pi/3$ and $5\pi/3$ respectively. Since $i\lambda =m^2$, it is easy to see by virtue of results such as (\ref{dD1}) and (\ref{ddSeff}) that the edges of the thermal windows are inflection points of the effective action of the model when the latter is considered as a function of $m$. 

\begin{itemize}
\item \underline{$q_3=0$ for $\a_*\neq 0$ }
\end{itemize}
In the middle of the above thermal windows we have  $\b\a_*=\pi$ (mod $2\pi$) and the imaginary charge vanishes $q_3=0$. Since $D_1(-1)=\ln 2$ we find that the thermal mass vanishes for ${\cal N}_3=-T\ln 2$. It is not a coincidence that this is {\it minus} that free energy of a massless free fermion in $1d$. 

\subsubsection{The free energy density}
The large-$N$ free energy density of the model has been calculated in  \cite{Filothodoros:2016txa}  as
\be
\label{freeEb}
\frac{1}{N}\D f^{(3)}_b(\b)\equiv \frac{1}{N}(f_b(\infty) -f_b(\b))=\frac{m_*^2}{g_{3,*}}+i\a_*q_3+\frac{1}{\p\b^3}D_3(z_*)\,,
\ee
where  $f(\b)=-\ln Z(\b)/(\b V_2)$. This also requires the subtraction of its linear divergence due to $1/g_{3,*}$, and then the physically relevant cases $q_3=0$ arise for $\b\a_*=0,\p$. In the first case the critical gap equation (\ref{gapb31}) has only the nonzero "golden mean"  solution for $m_*$, and one finds (see e.g. \cite{Sachdev:1993pr})
\be
\label{feCP0}
\frac{1}{N}\D f^{(3)}_b(\b)\Bigl|_{\b\a_*=0} =\frac{4}{5}\frac{\zeta(3)}{\pi \b^3}\,.
\ee
We can also calculate the free energy at the middle point of the bosonic thermal window when we tune ${\cal N}_3=-T\ln 2$ such that we have $m_*=0$, and we obtain
\be
\label{feCP0}
\frac{1}{N}\D f_b^{(3)}(\b)\Bigl|_{\b\a_*=\p} =-\frac{3}{4}\frac{\zeta(3)}{\pi \b^3}\,.
\ee
This is minus the free energy of a massless free Dirac fermion. Notice also that the free energy density of the {\it free} massless bosonic model corresponds to setting $g_3=0$ and $m_*=0$ and it is found to be\footnote{This is twice the free energy density of $N$ massless free scalars.}
\be
\label{feCPfree}
\frac{1}{N}\D f_{b,free}(\b) =\frac{\zeta(3)}{\pi \b^3}\,.
\ee
At the edges of the thermal windows we find
\begin{align}
\label{CPfeWindows2}
\frac{1}{N}\Delta f_{b}^{(3)}(\b)\Bigl|_{\b\a_*=\p/3}&=-\frac{1}{3\p\b^3}\left[\zeta(3)+ \pi Cl_2\left(\frac{\p}{3}\right)\right]\,,\\
\label{CPfeWindows2}
\frac{1}{N}\Delta f_{b}^{(3)}(\b)\Bigl|_{\b\a_*=5\p/3}&=-\frac{1}{3\p\b^3}\left[\zeta(3)- 5\pi Cl_2\left(\frac{\p}{3}\right)\right]\,.
\end{align}
Again, this appears to be the sum of contributions from conformal matter  (the $\zeta(3)$ term), and a hyperbolic volume  (the $Cl_2(\p/3)$ term).  

\subsection{The  $3d$ fermion-boson map}

The results in Sections 3.1 and 3.2 demonstrate the intimate relationship between the fermionic and the bosonic models. We can be more explicit and at the same time draw lessons for higher dimensions. Firstly, from the gap equations (\ref{GNgap1}), (\ref{gap31}) and  (\ref{CPgap1}), (\ref{gapb31}) we have, up to constant terms,
\begin{align}
\label{Sfeff}
S_{f,eff}&\sim -\frac{{\rm Tr}{\mathbb I}_2}{2\pi\b}\int \s[-{\cal M}_3\b+D_1(-z)]d\s\,,\\
\label{Sbeff}
S_{b,eff}&\sim \frac{2}{2\pi\b}\int [-{\cal N}_3\b+D_1(\hat{z})]mdm\,.
\end{align}
Notice the sign in (\ref{Sfeff}). From the above we see that if we make the identifications ${\cal M}_3 \leftrightarrow {\cal N}_3$ and $\s_*\leftrightarrow m_*$,  and the following shift of the imaginary chemical potential
\be
\label{gapmap}
\b\a_*\leftrightarrow \b\a_*+\p \,\,({\rm mod}\,2\p) \Rightarrow z_*\leftrightarrow -\hat{z}_* , \frac{2}{{\rm Tr}{\mathbb I}_2}Q_3\leftrightarrow -q_3\,,
\ee
then the free energies of the models are related to each other as
\be
\label{femap}
\frac{2}{N({\rm Tr}{\mathbb I}_{2})}\D f_f(\b)\Biggl|_{\b\a_*+ \pi} +\frac{1}{N}\D f_b(\b)\Biggl|_{\b\a_*}=\frac{1}{\b^3}D_2(z_*)=-i\frac{\p}{\b} q_3\,.
\ee
This explains the results in Sections 3.1.3 and 3.2.3. Notice that the divergent terms in the fermionic and the bosonic free energy densities come with the opposite sign and cancel in the sum. 

Relation (\ref{femap})  can be written as
\be
\label{pfduality3}
Z_{tot}(\b\a_*)\equiv Z^{(3)}_{f}(\b\a_*+\p)[Z^{(3)}_b(\b\a_*)]^{\frac{{\rm Tr}{\mathbb I}_{2}}{2}}=e^{i\pi V_2 N \frac{{\rm Tr}{\mathbb I}_{2}}{2}q_3}=e^{-N\frac{{\rm Tr}{\mathbb I}_{2}}{2}\frac{V_2}{\b^2}D_2(z_{*})}\,,
\ee
where $Z^{(3)}_f$ ($Z^{(3)}_b$) denote the $3d$ fermionic (bosonic) canonical partition function. We have kept the ${\rm Tr}{\mathbb I}_{2}$ explicitly in order to compare with the corresponding formula for general $d$ that will be given later. 
When $\b\a_*=0,\pi$, then $Q_3=q_3=0$ and (\ref{pfduality3}) is the well-known statement of fermion/boson duality i.e. the twisted fermionic (namely, imposing period boundary conditions) and the bosonic partition functions are inverse one of the other. However, for $Q_3,q_3\neq 0$ the corresponding partition functions are {\it weighted duals} due to the presence of the real exponential in the r.h.s. of (\ref{pfduality3}). We can give an interpretation of that latter weight factor recalling that the gap equations  (\ref{gap32}) and (\ref{gapb32}) tie $Q_3$  and $q_3$ to the Bloch-Wigner function $D_2(z)$, and hence to the  volume of hyperbolic manifolds. Then (\ref{pfduality3}) could be understood as giving the leading "classical" term in a perturbative expansion of a complex Chern-Simons action in inverse powers of the level \cite{Gukov:2016njj}. Support for such an interpretation also comes from the fact that the extremal values that we have found for the fermionic and bosonic imaginary chagres, coincide with the results reported e.g. in \cite{Gukov:2016njj,Gang:2017hbs} in the study of the partition function of the $SL(2,\mathbb{C})$ CS theory. 


\section{Towards the femion-boson map for all odd $d>3$}

Having exhaustively reviewed the phase structure of the $3d$ models, lets go back to a point that we have raised in the Section 2. In the context of statistical physics the imaginary chemical potential is simply the variable  generating the Legendre transform to the canonical ensemble. We have seen, however,  that its introduction touches upon the microscopic statistics of the system since on the one hand it unveils  the existence of "statistical transmutation thermal windows", and on the other hand its dimensionality. For example, 
in the $3d$ above we noticed that even without a detailed knowledge of the underlying microscopic system, we could have made  an educated guess about some of its properties. Namely, we could conclude that the microscopic system would be bosonic if, for a given chemical potential $\a_*\neq 0$, the edge of its first thermal window is at a higher temperature.  Moreover, by observing that this temperature is  {\it twice} as large as the corresponding fermionic value,  we could also conclude  that the microscopic system is three dimensional. The result would then be confirmed by the pattern of the positions of all the thermal window edges. Does this or a similar picture exists in higher dimensions? This leads us to consider the generalization of the three-dimensional models to $d>3$. At this point we observe that the generalization of the Gross-Neveu and CP$^{N-1}$ models to  {\it even} $d>3$ is rather complicated and presumably unrelated to the physics that we want to describe. For example, it is well-known (e.g. \cite{Brandt:2012mu}) that thermal field theories have different analyticity properties with respect to their couplings in $d$ even and odd dimensions. This is also apparent in the calculation of the free energy density of massive free scalars and fermions which for $d$-odd can be expressed as finite sums of Nielsen's generalized polylogarithms \cite{Filothodoros:2016txa, Borwein,Kolbig}, while for $d$-even the corresponding expressions are much more complicated. Moreover, we have already noted the relevance of the 1$d$ theories to the physics of the the 3$d$ models. For these reasons, we will concentrate on studying the generalizations of our 3$d$ models to {\it odd} $d>3$.

\subsection{The fermions}
The GN model in $d$ Euclidean dimensions is described by the generalization of the   3$d$ action (\ref{GNaction})
\begin{equation}
\label{GNdaction}
S_{GN} = -\int_0^\b \!\!\!dx^0\int \!\!d^{d-1}\bar{x} \left[\bar{\psi }^{a}(\slash\!\!\!\partial  -i\gamma_0\alpha)\psi ^{a}+\frac{G_d}{2({\rm Tr}\mathbb I_{d-1})N}\left (\bar{\psi }^{a}\psi ^{a}\right )^{2} +i\a NQ_d\right]\,,
\end{equation}
with  $Q_d$ the $N$-normalized $d$-dimensional fermionic number density and $a=1,2,..N$.  For odd $d$ we take the dimension of the gamma matrices to be ${\rm Tr}{\mathbb I}_{d-1}=2^{\frac{d-1}{2}}$.


The $d$-dimensional  gap equations become
\begin{align}
\label{GNdgap1}
\frac{\sigma_*}{G_d}&=\frac{\sigma_*}{\b}\sum_{n=-\infty}^\infty\int^\Lambda\!\!\frac{d^{d-1} \bar{p}}{(2\pi)^{d-1}}\frac{1}{\bar{p}^2+(\omega_n-\alpha_*)^2+\sigma_*^2}\,,\\
\label{GNdgap2}
i Q_d&=\lim_{\epsilon\rightarrow 0}\frac{{\rm Tr}\mathbb I_{d-1}}{\b}\int^\Lambda\!\!\frac{d^{d-1} \bar{p}}{(2\pi)^{d-1}}\sum_{n=-\infty}^\infty\frac{e^{i\omega_n\epsilon}(\omega_n-\alpha_*)}{\bar{p}^2+(\omega_n-\alpha_*)^2+\sigma_*^2}\,.
\end{align}
The main issue with the GN model in $d>3$ is that the gap equation (\ref{GNdgap1}) has  a finite number of higher order divergent terms as $\Lambda\rightarrow\infty$, which cannot be simply taken care of by the adjustment/renormalization of the single coupling $G_d$.  On the other hand the charge gap equation (\ref{GNdgap2}) is  cut-off independent. Despite these obstructions we will be able to extract useful information regarding the phase structure of the model, albeit not as clear cut as in $d=3$. In particular we will exhibit the generalization of the three-dimensional fermion-boson map.

We will discuss below in some detail the cases $d=5$ and $d=7$ in order to exhibit some of the general features of the higher dimensional models. Starting with $d=5$ and using the results given in the Appendix we find the form of two gap equations as
\begin{align}
\label{gap51}
\s_*\left[-{\cal M}_5\b^3-D_3(-z_*)-\frac{1}{2}\ln^2\!|z_*|\left(D_1(-z_*)-\frac{2}{3\pi}\gamma\right)\right]&=0\,,\\
\label{gap52}
\frac{(2\pi)^2}{{\rm Tr}\mathbb I_{4}}\b^4Q_5-3i\left[D_4(-z_*)+\frac{1}{6}\ln^2\!|z_*|D_2(-z_*)\right]&=0\,,
\end{align}
where 
\be
\label{M5gamma}
\frac{{\cal M}_5}{(2\pi)^2}=\frac{1}{G_{5,*}}-\frac{1}{G_5}\,,\,\,\,\,\gamma=\Lambda\beta\,.
\ee
To derive (\ref{gap51}) we have dropped the infinite number of terms that go as inverse powers of $\Lambda$ and we note that the last term in parenthesis resembles the corresponding three-dimensional gap equation (\ref{gap31}). Also, in (\ref{gap52}) we could have used (\ref{gap32}) to write it in terms of the charge $Q_3$ of a three-dimensional fermionic model. We thus begin to see signs of a partial deconstruction of the higher dimensional models in terms of lowers dimensional quantities. 

As we go to dimensions $d>3$, the crucial issue is the explicit presence of  the cutoff in the gap equation i.e. compare (\ref{gap51}) with (\ref{gap31}). We emphasise that ${\cal M}_5$ is independent of the cutoff $\Lambda$, hence for a given temperature (\ref{gap51}) is a two-parameter equation for $z_*$. This means that there is no unambiguous way to tune the single coupling constant of the theory, namely the parameter ${\cal M}_5$, in order to obtain a cutoff independent result. 
The charge gap equation on the other hand has no such issues.

 Since we eventually want to send $\Lambda\rightarrow\infty$, meaningful results for generic values of $z_*$ could be obtained in a high temperature regime where the parameter $\gamma>0$ remains finite. To begin our analysis we notice that for zero chemical potential $\a_*=0$ we can tune $\gamma$ such that the critical critical gap equation ${\cal M}_5=0$  always has a non-zero solution $\sigma_*<0$. This solution disappears if we also tune ${\cal M}_5$ to a suitable positive value i.e. if we enter the strong coupling regime. However, this observation is not universal as it involves the arbitrary fine tuning of two parameters. More robust is the study of the gap equation  for $\sigma_*=0$. Indeed, taking a second derivative wrt $\s$ of (\ref{gap51}), which is equivalent to taking the second derivative of the five-dimensional effective action, we find
 \be
 \label{ddSeff5}
\frac{\partial^2}{\partial\sigma^2}S^{(5)}_{f,eff}\Biggl|_{\substack {\sigma_*=0}}\propto-{\cal M}_5\b^3+D_3(-e^{-i\b\a_*})\,.
\ee
Then, for $\a_*=0$ the special value 
\be
\label{M5crit}
{\cal M}_5=-D_3(-1)T^3=\frac{3}{4}\zeta(3)T^3
\ee
corresponds to the condition that $\s_*=0$ is an inflection point of the effective action. Up to a geometric factor, this is the free energy of massless free Dirac fermions in $d=3$ i.e. (\ref{freeEf0}). In Appendix A we give the formulae for the free energy densities of free fermionic (\ref{Dfd}) and bosonic (\ref{Dbd}) theories in $d$ dimensions 
and we note that (\ref{M5crit}) is a special case of the general result
\be
\label{Mdcrit}
{\cal M}_d=\frac{S_d}{2{\rm Tr}\mathbb I_{d-1}}T^d\Delta f_{f}^{(d-2)}(\b)\Bigl|_{m=0}\,,
\ee
for $d=3,5,7,..$

It is interesting then to note that the $D_d(e^{-x})$ functions, for odd $d$, behave near $x=0$ as
\be
\label{Ddx0}
D_d(\pm e^{-x})=D_d(\pm1)+c^{\pm}(d) x^{d+1}+O(x^{d+3})\,,
\ee
with $c^{\pm}(d)$ numerical factors. This appears to be pointing at a possible mechanism for fixing $\gamma$ in (\ref{gap51}). Namely, by tuning $\gamma=3\pi(\ln 2)/2$ we would arrange that the effective action behaves near $\s_*=0$ as $S_{f,eff}^{(5)}\sim \b^2V_4\s^6$, with $V_4$ the 4$d$ spatial volume.  This is reminiscent of the fine tuning needed to obtain multicritical fixed points see e.g. \cite{Itzykson:1989sx}. Nevertheless, as we will see below such a procedure does not work for $d>5$ as the additional parameters that enter the gap equation are not independent but rather come as powers of $\gamma$.  This limitation comes from the simplicity of our models. It would be interesting to see whether there are more elaborate models on which the above tuning procedure could be consistently applied.

In the presence of an imaginary chemical potential the situation becomes more interesting, since we encounter again nontrivial zeros of $D_3(-z_*)$ on the unit circle. That means that we can now study the critical theory with ${\cal M}_5=0$. A short excursion in Mathematica  yields two zeroes for $D_3(-z)$ on the unit circle. Remarkably their positions are approximated to high accuracy by rational multiples of $\pi$ as
\be
\label{Cl3}
D_3(-e^{-i\b\a_*})=Cl_3(\b\a_*\pm\p)=0\Rightarrow \b\a_*\approx \frac{7\p}{13}\,{\rm or}\,\b\a_*=\frac{19\p}{13} \,\,\,({\rm mod}\,2\p)\,.
\ee
Using the periodic properties of the Clausen functions, the relevant results are 
\be
\label{Clausen5}
Cl_3\left(\frac{6\p}{13}\right)=Cl_3\left(\frac{20\p}{13}\right)=0.000362159\,.
\ee
From  (\ref{DmDm-1}) and by virtue of (\ref{gap52}) we find at these points 
\be
\label{Q5max}
Q_{5,extr}=\pm i{\rm Tr}\mathbb I_{4}\frac{2}{S_5\b^4}Cl_4\left(\frac{6\p}{13}\right)\,, \,\,\,\,\,S_5=\frac{8\pi^2}{3}
\ee
since $Cl_4(\pm 6\p/13)\approx \pm 0.995777$ are the maximum (minimum) values of $D_4(-z)$ on the unit circle. Notice that $S_5\b^4$ is the surface of the 4-dimensional sphere. We will see that this patterns generalises to all dimensions.

Finally, when $\b\a_*=\p$  the gap equation (\ref{gap51}) coincides - apart the overall $\s_*$ factor - with the corresponding one of the $CP^{N-1}$ that will be given below. The charge is $Q_5=0$ and the system has been bosonized. However, in contrast with the analogous situation in $d=3$, a nonzero solution for $\s_*$ in the critical case ${\cal M}_5=0$ depends on the arbitrary parameter $\gamma$. 



We then briefly discuss the seven-dimensional case which shows how our results generalize to higher dimensions. The gap equations are
\begin{align}
\label{gap71}
&\s_*\left[-{\cal M}_7\b^5+D_5(-z_*)+\frac{1}{6}\ln^2\!|z_*|\left(D_3(-z_*)+\frac{\gamma^3}{45\p}\right)+\right.\nonumber \\
&\hspace{6cm}\left.+\frac{1}{24}\ln^4\!|z_*|\left(D_1(-z_*)-\frac{4\g}{15\p}\right)\right]=0\,,\\
\label{gap72}
&\frac{(2\pi)^3}{{\rm Tr}\mathbb I_{6}}\b^6Q_7+15i\left[D_6(-z_*)+\frac{1}{10}\ln^2\!|z_*|D_4(-z_*)+\frac{1}{120}\ln^4\!|z_*|D_2(-z_*)\right]=0\,,
\end{align}
where the parameter $\g$ has been defined above, and 
\be
\label{M7}
\frac{3{\cal M}_7}{(2\pi)^3}=\frac{1}{G_{7,*}}-\frac{1}{G_7}\,.
\ee
As before, we clearly see in (\ref{gap71}) the presence of terms related to the corresponding three- and five-dimensional  gap equations (\ref{gap31}) and (\ref{gap51}), as well as the appearance of the charges $Q_3$ and $Q_5$, through $D_2(-z)$ and $D_4(-z)$,  in (\ref{gap72}). The condition that $\s_*=0$ is an inflection point of the effective action is now can 
\be
\label{M7crit}
{\cal M}_7=D_5(-1)T^5=\frac{15}{16}\zeta(5)T^5\,,
\ee
and follows the general formulae (\ref{Dfd}) and (\ref{Mdcrit}). Moreover as advertised above we see that it is not possible tuning $\gamma$ to remove the constant terms $D_3(-1)$ and $D_1(-1)$ in the expansion of  (\ref{gap71}) near $\s_*=0$, and hence to arrange unambiguously for multicritical behaviour for the effective action. This problem will clearly persist for all $d>7$. 

Moving on the non zero chemical potential we can look for zeros of the critical gap equation (\ref{gap72})  on the unit circle. Again, their positions are remarkably well approximated, better than in $d=5$, by rational multiples of $\pi$ as
\be
\label{Cl5}
D_5(e^{-i\b\a_*})=Cl_5(\b\a_*\pm \p)=0\Rightarrow \b\a_*\approx \frac{26\p}{51}\,{\rm or}\,\frac{76\p}{51} \,\,\,({\rm mod}\,2\p)\,.
\ee
The relevant result is 
\be
\label{Clausen7}
Cl_5\left(\frac{25\p}{51}\right)=Cl_5\left(\frac{77\p}{51}\right)=0.000129657\,.
\ee
Using then (\ref{DmDm-1}) we find at these points by virtue of (\ref{gap52}) 
\be
\label{Q5max}
Q_{7,extr}=\mp i{\rm Tr}\mathbb I_{6}\frac{2}{S_7\b^6}Cl_6\left(\frac{25\p}{51}\right)\,,\,\,\,\,\,S_7=\frac{16\pi^3}{15}\,,
\ee
since $Cl_6(\pm 25\p/51)\approx \pm 0.999151$ are  the maximum (minimum) values of $D_6(-z)$ on the unit circle.

The basic features discussed above do not change as we move to higher dimensions. We continue to see the partial deconstruction of the $d$-dimensional gap equations in terms of lower-dimensional pieces. Namely, (\ref{GNdgap1}) contains the $d-2, d-4,...,5,3$-dimensional gap equations,  and (\ref{GNdgap2}) contains  the $Q_{d-2},Q_{d-4},...,Q_5,Q_3$ charges. The parameter $\gamma$ appears in the form of an odd polynomial of degree $d-4$, and the condition that $\s_*=0$ is an inflection point of the effective action is  (\ref{Mdcrit}). 

\subsection{Spinoff: the zeros and the extrema of Clausen's functions}
More intriguingly, we seem to be able to give an analytic formula for the approximate positions of the zeroes of all $D_{2n-1}(z)$, $n=1,2,..$  functions on the unit circle. We obtain
\begin{align}
\label{Conjecture1}
&D_{2n-1}(e^{-i\b\a_*})\equiv Cl_{2n-1}(\b\a_*)=0\Leftrightarrow \b\a_{*}\approx \theta_n,2\pi -\theta_n\,({\rm mod}\,2\p)\,,
\\
\label{ThetaN}
&\theta_n=\frac{\pi}{2}\left(1-\frac{5}{4^{n+1}-(-1)^{n+1}}\right)\,,
\end{align}
for $n=1,2,3,..$. Using the periodic properties of Clausen functions, these determine the zeros of $Cl_{2n-1}(-z)$ on the unit circle  as well. At these points therefore the even Clausen functions $Cl_{2n}(e^{-i\b\a_*})$, and hence the corresponding imaginary charges of the models, obtain their maximum/minimum values respectively. Table 1. gives a small list of values for $Cl_{2n-1}(\theta_n)$ and $Cl_{2n}(\theta_n)$. 
We notice the extremely fast convergence in the accuracy of our formula (\ref{ThetaN}), and the fact that the maximum of $Cl_{2n}(\theta)$ approaches 1 very rapidly.  The result for  $n\rightarrow \infty$ follows from the  observation that this limit is actually well defined
\be
\label{ClausenInfinity}
\lim_{n\rightarrow\infty}Cl_{2n-1}(\theta)=\cos\theta\,,\,\,\,\lim_{n\rightarrow\infty}Cl_{2n}(\theta)=\sin\theta\,.
\ee

\begin{tabular}{ |p{3cm}||p{3cm}|p{3cm}|p{3cm}|  }
 \hline
 \multicolumn{4}{|c|}{{\bf Table 1}. The approximate zeroes of $Cl_{2n-1}(\theta)$ and corresponding maxima of $Cl_{2n}(\theta)$} \\
 \hline
$n$& $\theta_n$ &$Cl_{2n-1}(\theta_n)$&$Cl_{2n}(\theta_n)$\\
 \hline
 1  & $\frac{\p}{3}$    & 0&  1.01494\\
2&   $\frac{6\pi}{13}$  & 0.000362159  &0.995777\\
 3 &$\frac{25\pi}{51}$ & 0.000129657&  0.999151\\
4  &$\frac{102\pi}{205}$ & -0.000101475&  0.99988\\
 5 &   $\frac{409\pi}{819}$ & -0.0000317333&0.999985\\
 6& $\frac{1638\pi}{3277}$  & -8.71361$\times 10^{-6}$   &0.999998\\
 ..& ..  & ..   &..\\
$\infty$ &$ \frac{\pi}{2} $ & 0&1\\
 \hline
\end{tabular}

\begin{figure}[!htb]
\centering
\includegraphics[scale=1.2]{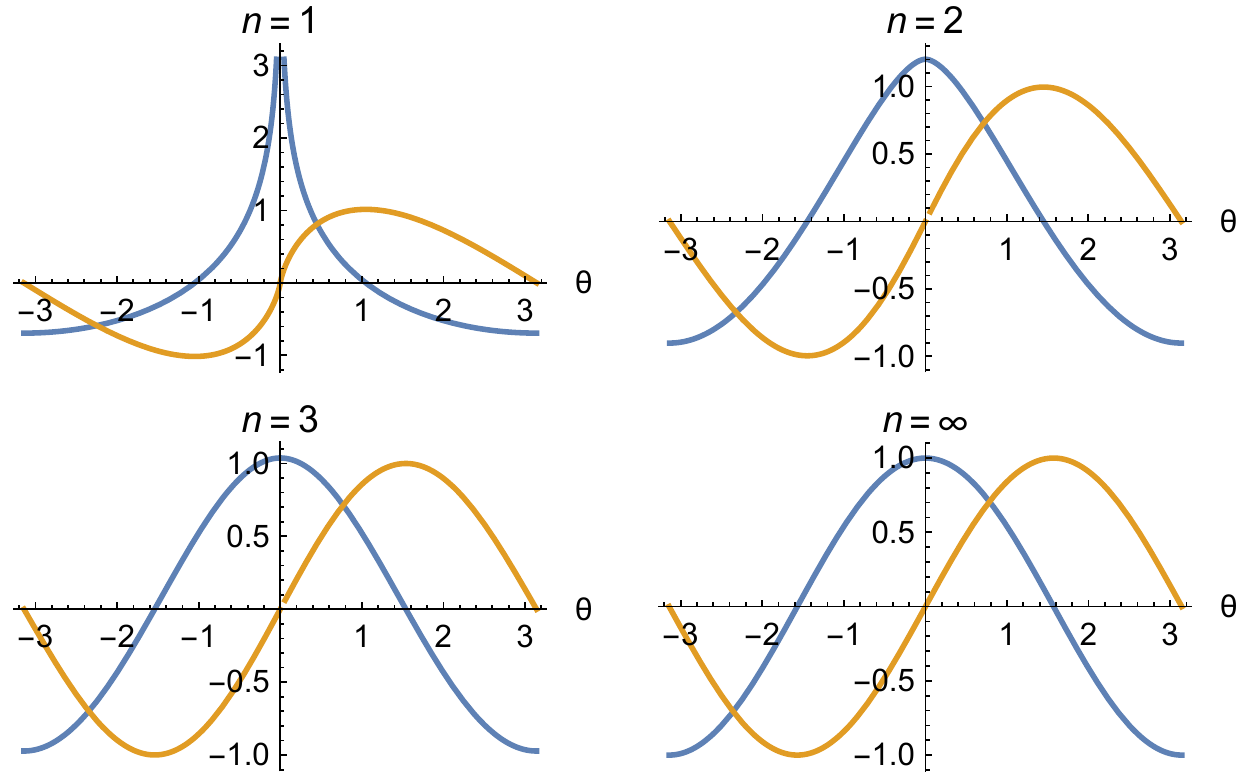}
\caption{The functions $Cl_{2n-1}(\theta)$ (blue line) and $Cl_{2n}(\theta)$ (yellow line) for $\theta\in[-\pi,\pi]$. }
\label{fig:digraph}
\end{figure}

\subsection{The bosons}
The action is the natural extension of (\ref{CPaction}) for general $d$
\be
\label{CPactiond}
S_{CPN}=\int_0^\b \!\!\!dx^0 \!\!\int \!\!d^dx\left[|(\partial_0-i\a)\phi^a|^2 +|\partial_i\phi|^2 +i\lambda(\bar{\phi}^a\phi^a-\frac{N}{g_d})+iNq_d\alpha\right]\,,\,\,\,a=1,2,..,N\,,
\ee
With the experience gathered from the three-dimensional case, it is not surprising to find the the bosonic formulae for the gap equations and the free energy can be obtained from the corresponding fermionic ones by the identification of the saddle points i.e. $\s_*=m_*$ and the shift $z_*\leftrightarrow -z_*$. For example, the gap equations in $d=5$ are
\begin{align}
\label{CPgap51}
&-{\cal N}_5\beta^3-D_3\left( z_*\right)-\frac{1}{2}\ln^2|z_*|\left(D_{1}\left( z^*\right)-\frac{2\g}{3\pi}\right)=0\,, \\
\label{CPgap52}
&\left( 2\pi\right)^2\beta^4q_5+3i\left[D_4\left(z_*\right)+\frac{1}{6}\ln^2|z^*|D_2\left(z_*\right)\right]=0\,,
\end{align}
and the parameter ${\cal N}_5$ by the bosonic version of (\ref{M5gamma}). It is clear that the discussion regarding the phase structure of the bosonic models will be the shifted image of the corresponding fermionic ones, and we therefore spare the reader from repeating it. The higher dimensional results also follow the same pattern.

\subsection{The fermions-bosons map for large-$d$}
The large-$N$ free energy density of the Gross-Neveu model for general odd $d$ is given by 
\begin{align}
\label{GNfed}
\frac{1}{N{\rm Tr}{\mathbb I}_{d-1}}\D f^{(d)}_{f}(\b) =& \frac{1}{2}\int \frac{d^dp}{(2\pi)^d}\ln\left(\frac{p^2+\s_*^2}{p^2}\right)+\frac{1}{\b}\int\frac{d^{d-1}\bar{p}}{(2\pi)^{d-1}}\Re \left[\ln\left(1+e^{-\b\sqrt{\bar{p}^2+\s_*^2}-i\b\a_*}\right)\right]\nonumber \\
&-\frac{1}{2}\int^{\Lambda}\!\!\frac{d^dp}{(2\pi)^d}\frac{\s_*^2}{p^2+\s_*^2}-\a_*\Im[j_d(\sigma_*,\a_*)]+\frac{1}{2}\s_*^2\Re[i_d(\s_*,\a_*)]\,.
\end{align}
The integrals $i_d(\s_*,\a_*)$ and $j_{d}(\s_*,\a_*)$ are defined in (\ref{id}) and (\ref{jd}) respectively. As advertised, due to the first term in the second line of (\ref{GNfed}) and by virtue of (\ref{zeroTgap}) the free energy contains a finite number of divergent terms that appear  as an odd polynomial of degree $d-4$ in the parameter $\g=\Lambda\b$. However, we find it remarkable that the $\Lambda$-independent terms conspire in such a way as to enable us to express (\ref{GNfed}) as a linear combination of the $D_d(-z)$ functions of both odd and even $d$, with the latter contributions coming exclusively from the $j_d(\s_*,\a_*)$ term. For example, using some useful formulae presented in Appendix Δ we obtain for $d=5$ 
\begin{align}
\label{GNfe5}
\frac{1}{N{\rm Tr}{\mathbb I}_{4}}\D f_f^{(5)} =&-\frac{\s_*^2}{2G_{5,*}}+\frac{\b^4\s_*^4\gamma}{24\pi^3 \b^5}+\frac{3}{4\pi^2\b^5}\left[D_5(-z_*)-\frac{1}{24}\ln^4|z_*|D_1(-z_*)\right]\nonumber\\
&+\frac{\b\a_*}{8\pi^2\b^5}\left[D_4(-z_*)+\frac{1}{6}\ln^2|z_*|D_2(-z_*)\right]\,.
\end{align}
The presence of the two divergent terms in the first line make it hard to trust any value of the free energy for $\s_*\neq 0$. When $\s_*=0$ we either take $\a_*=0$ when we obtain the free theory result (\ref{Dfd}), or we consider the edges of the generalized thermal windows $\b\a_*=7\pi/13$ where we obtain 
\be
\label{GNfe5Windows}
\frac{1}{N{\rm Tr}{\mathbb I}_{4}}\D f_f^{(5)}\Bigl|_{\b\a_*=7\pi/13} =\frac{3}{4\pi^2\b^5}D_5(-e^{-i\frac{7\pi}{13}})+\frac{7}{104\p\b^5}D_4(-e^{-i\frac{7\pi}{13}})\,,
\ee
and an equivalent result for $\b\a_*=19\pi/13$. Unfortunately $D_5(-e^{-i7\pi/13})$ does not seem to be related to a rational multiple of $\zeta(5)$. This pattern  continues in higher dimensions, namely the Bloch-Wigner-Ramakrishnan functions appear to be the suitable basis for the expansion of the finite part of the free energy density of fermionic systems in $d>5$. 

The large-$N$ free energy of the CP$^{N-1}$ model for general $d$ is given by
\begin{align}
\label{CPfed}
\frac{1}{N}\D f^{(d)}_{b}(\b) =& -\int \frac{d^dp}{(2\pi)^d}\ln\left(\frac{p^2+\s_*^2}{p^2}\right)-\frac{2}{\b}\int\frac{d^{d-1}\bar{p}}{(2\pi)^{d-1}}\Re \left[\ln\left(1-e^{-\b\sqrt{\bar{p}^2+\s_*^2}-i\b\a_*}\right)\right]\nonumber \\
&+\int^{\Lambda}\!\!\frac{d^dp}{(2\pi)^d}\frac{\s_*^2}{p^2+\s_*^2}+2\a_*\Im[j_d(\sigma_*,\a_*+\frac{\pi}{b})]-\s_*^2\Re[i_d(\s_*,\a_*+\frac{\pi}{b})]\,.
\end{align}
Notice the overall factor of 2 and the fact that the divergent terms come with the opposite sign compared to the fermionic case. The latter is a crucial observation that will allow us to define the fermion-boson map in an unambiguous way for general $d$. 

The $d=5$ free energy is then given by
\begin{align}
\label{CPfe5}
\frac{1}{N}\D f_b^{(5)} =&+\frac{\s_*^2}{g_{5,*}}-\frac{\b^4\s_*^4\gamma}{12\pi^3 \b^5}-\frac{3}{2\pi^2\b^5}\left[D_5(z_*)-\frac{1}{24}\ln^4|z_*|D_1(z_*)\right]\nonumber\\
&-\frac{\b\a_*}{4\pi^2\b^5}\left[D_4(z_*)+\frac{1}{6}\ln^2|z_*|D_2(z_*)\right]\,.
\end{align}
Then, the $d=5$ fermionic and bosonic free energy densities are related as
\be
\label{feduality5}
\frac{2}{N{\rm Tr}{\mathbb I}_{4}}\D f_f^{(5)}(\b)\Bigl|_{\b\a_*+\p}+\frac{1}{N}\D f_b^{(5)}\Bigl|_{\b\a_*}=i\frac{\pi}{6\b}q_5\,.
\ee
The corresponding result for the partition functions is
\be
\label{pfduality5}
Z_{tot}^{(5)}(\b\a_*)\equiv Z_f^{(5)}(\b\a_*+\pi)\left[Z_{b}^{(5)}(\b\a_*)\right]^{\frac{{\rm Tr}{\mathbb I}_{4}}{2}}=e^{-iNV_4\frac{\pi}{6}q_5}\,.
\ee
To generalize (\ref{pfduality5}) for all $d$ we note firstly that the exponent ${\rm Tr}{\mathbb I}_{d-1}/2$  of the bosonic partition function is a manifestation of the usual fact that as we go up in dimension a single Dirac fermion "weights" more and more bosonic degrees of freedom. Hence, if we choose a model with a supersymmetric matter content,\footnote{We do not want imply that there is some sort of  supersymmetry here, as this would require many more additional d.o.f.} such that the number of bosons is $2^{\frac{d-1}{2}}$, then we could get rid of that $d$-dependent exponent. Next, we observe that the contribution on the r.h.s. of (\ref{feduality5}), or equivalently the exponential in (\ref{pfduality5}), comes from the $\Im[j_d(\s_*,\a+\pi/b)]$ term in the shifted fermionic free energy (\ref{CPfed}). This integral is given in (\ref{jdNielsen}) in terms of Clausen functions. Putting all these together, we can give the formula for the partition function duality between fermionic and bosonic theories, with supersymmetric matter content, at $\s_*=0$ and for general $d$ as
\be
\label{pfdualityd0}
Z^{(d)}_{"SUSY"}(\b\a_*)\Bigl|_{\s_*=0}\equiv\left[Z_f^{(d)}(\b\a_*+\pi)Z_{b}^{(d)}(\b\a_*)\right]\Bigl|_{\s_*=0}=e^{4\pi N\frac{V_{d-1}}{S_d \b^{d-1}}Cl_{d-1}(\b\a_*)}
\ee
The result (\ref{pfdualityd0}) suggests the existence of a non trivial large-$d$ limit of the fermion-boson duality. Namely, if we take the zero temperature or decompactification limit $\b\rightarrow\infty$ the ratio $V_{d-1}/S_{d}\b^{d-1}\rightarrow 1$ (i.e. we can think of $V_{d-1}$ as the surface of a very large sphere). Moreover, the Clausen's functions with even index, such as those that appear in (\ref{pfdualityd0}), have a well defined $d\rightarrow\infty$ limit which is simply $\sin(\b\a_*)$. Therefore we can write
\be
\label{pfsusylimit}
\lim_{d\rightarrow\infty} Z^{(d)}_{"SUSY"}(\b\a_*)\Bigl|_{\s_*=0}=e^{4\pi N\sin(\b\a_*)}\,.
\ee
It would be interesting to understand this formula better.


\section{Summary and discussion}

The main message of our work is that the well-known 3$d$ map between fermions and bosons generalises to all odd $d$ dimensions and it is unveiled by the presence of an imaginary chemical potential for a $U(1)$ global charge. In three-dimensions a detailed analysis shows explicitly the map between the phase structure of the fermionic and bosonic models. A similar analysis in higher dimensions is not possible due to the non renormalizability of the corresponding models, nevetheless a precise map of their gap equations, free energies and partition functions can be demonstrated for a certain region of their phase space. 

Our calculations have unveiled the relevance of the Bloch-Wigner-Ramakrishnan functions $D_d(z)$ to the physics of the fermion-boson map. Firstly, it appears that the $D_d(z)$'s are the natural functional basis for the expansion of the gap equations and the free energies of the models. Then, the zeros and extrema of these functions are saddle points of the systems. Also, since $D_2(z)$ expresses an hyperbolic volume, our 3$d$ results are related to recent studies of complex non-abelian Chern-Simons theories, and our higher dimensional ones provide possible generalisations. Finally, we have argued that there is a non-trivial large $d$ limit of the fermion-boson map which at the level of partition functions is expressed by the formula (\ref{pfsusylimit}). 

We think that our results offer a new window into the physics of bosonisation. As we briefly alluded to in the text, for a statistical physics system it appears that bosonisation is intimately related with a Legendre transformation to the canonical ensemble - but for purely imaginary global charge. In such a context and in the absence of a microscopic model one needs some other parameter to uncover the statistical properties of the elementary d.o.f. The dimensionality $d$, which e.g. can be taken to be the number of nearest neighbours in a lattice model \cite{Georges:1996zz}, could be a useful parameter if a suitable $1/d$ expansion can be implemented.  It would also be interesting to study the partition functions on higher spheres extending the results in \cite{Giombi:2017txg}. 

The purely mathematical aspect of our results remains a mystery to us. It would be nice to have a better understanding of our approximation formula (\ref{ThetaN}) for the zeros and exrtema of the Clausen functions. It would also be interesting to explore whether  $D_d(z)$ functions  with even $d$ are related to higher hyperbolic volumes. Finally, the precise relationship of our results with the wealth of recent work on resurgence of complex Chern-Simons theory, is another intriguing and open question.

\section*{Acknowledgements}
ACP would like to thank the Theoretical Physics Department of CERN for its warm hospitality, where the final part of this work was completed. He would also like to acknowledge useful correspondence with D. Gang and M. Mari\~{n}o.

\appendix

\section{Notation and useful results}
Our conventions are $x^{\m}=(x^0,x^i)$, $\m,\n,..=0,1,2,..d-1$, $i,j,..= 1,2,..d-1$, $\bar{x}=(x^i)$, and analogously for the momentum $p^{\m}$. Our spinor notation follows \cite{ZinnJustin:2002ru}. For example, in $d=3$ we use two-component Euclidean Dirac spinors and a Hermitian representation for the gamma matrices as $\gamma_\m=\s_\m$ where $\m=0,1,2$. $\sigma_\m$ are the usual Pauli matrices with the definition $\gamma_0=\sigma_0\equiv \sigma_3$.  Latin indices run as $i=1,2$. For odd dimensions $d>3$ we use gamma matrices with dimension ${\rm Tr}{\mathbb I}_{d-1}=2^{\frac{d-1}{2}}$.

We perform the Matsubara sums using the Poisson sum formula
\be
\label{Poisson}
\sum_{n=-\infty}^\infty f(n)=\sum_{k=-\infty}^\infty\int_{-\infty}^\infty dx \,e^{-i2\pi kx}f(k)\,.
\ee
For the fermionic gap equations we have $\omega_n=\p(2n+1)/\b$ and we need the formula
\begin{align}
\label{gapexample1}
&\frac{1}{\b}\sum_{n=-\infty}^\infty\int^\Lambda\!\!\!\frac{d^{d-1} p}{(2\pi)^{d-1}}\frac{1}{\bp^2+(\omega_n-\alpha_*)^2+\s_*^2}  = \int^\Lambda\!\!\!\frac{d^d p}{(2\pi)^d}\frac{1}{p^2+\s_*^2}-\Re \,[i_{d}(\sigma_*,\a_*)]\,,\\
\label{idfer}
&i_{d}(\sigma_*,\a_*)=\int \frac{d^{d-1}\bp}{(2\pi)^{d-1}}\frac{1}{\sqrt{\bp^2+\s_*^2}}\frac{1}{1+e^{\b\sqrt{\bp^2+\s^2_*}+i\b\a_*}}\,.
\end{align}
For the bosonic gap equations we have $\omega_n=2n\pi/\b$ and we need instead
\be
\label{gapexample2}
\frac{1}{\b}\sum_{n=-\infty}^\infty\int^\Lambda\!\!\!\frac{d^{d-1} p}{(2\pi)^{d-1}}\frac{1}{\bp^2+(\omega_n-\alpha_*)^2+\s_*^2}  = \int^\Lambda\!\!\!\frac{d^d p}{(2\pi)^d}\frac{1}{p^2+\s_*^2}-\Re \,[i_{d}(\sigma_*,\a_*+\frac{\pi}{\b})]\,,
\ee
The cutoff dependance is not relevant for $i_d(\sigma_*,\a_*)$ as the integral is finite. A standard inversion formula for the hypergeometric function allows us to obtain \begin{align}
\label{zeroTgap}
\int^\Lambda\!\!\!\frac{d^d p}{(2\pi)^d}\frac{1}{p^2+\s_*^2} &= \frac{1}{G_{d,*}}-\frac{\Lambda^{d-2}}{d-2}\frac{S_d}{(2\pi)^d}{}_2F_1\left(1,\frac{d}{2}-1;\frac{d}{2};-\frac{\Lambda^2}{\sigma_*^2}\right)\nonumber\\
&\hspace{-2cm}=\frac{1}{G_{d,*}}-\frac{S_d}{(2\pi)^d}\left[\Gamma\left(\frac{d}{2}\right)\Gamma\left(2-\frac{d}{2}\right)\frac{\sigma_*^{d-2}}{d-2}+\frac{\sigma^2_*\Lambda^{d-4}}{d-4}{}_2F_1\left(1,2-\frac{d}{2};3-\frac{d}{2};-\frac{\sigma_*^2}{\Lambda^2}\right)\right]
\end{align}
where $S_d=2\pi^{d/2}/\Gamma(d/2)$. This way we  see that for odd $d$ there are a finite number of divergent terms as $\Lambda\rightarrow\infty$. To evaluate $i_d(\sigma_*,\a_*)$ we set $z=e^{-\b\sqrt{\bp^2+\sigma_*^2}-i\b\a_*}$ and we obtain
\be
\label{id}
i_{d}(\sigma_*,\a_*)=\frac{S_{d-1}}{(2\pi)^{d-1}}\frac{1}{\b^{d-2}}\int_0^{z_*}\frac{dz}{z+1}\left[(\ln z+i\b\a_*)^2-\b^2\s_*^2\right]^{\frac{d-3}{2}}\,,\,\,\,z_{*}=e^{-\b\sigma_*-i\b\a_*}
\ee
which eable us to calculate
\be
\label{Reid}
\Re[i_{d}(\sigma_*,\a_*)]=\frac{1}{2}[i_{d,f}(\sigma_*,\a_*)+\bar{i}_{d,f}(\sigma_*,\a_*)]
\ee
as a sum of $D_d(z)$ functions with odd index.
For $\s_*=0$ the above integral can be easily evaluated using as an intermediate step its representation as a Nielsen generalised polylogarithm $S_{n,p}(z)$ \cite{Kolbig,Borwein} 
\be
\label{Nielsen}
S_{n,p}(z)=\frac{(-1)^{n+p-1}}{(n-1)!p!}\int_0^1\ln^{n-1}(x)\ln^p(1-zx)\frac{dx}{x}\,,\,\,\,n,p=1,2,3,..\,, 
\ee
and then using $S_{n,1}(z)=Li_{n+1}(z)$ we finally find
\be
\label{idNielsen}
i_d(0,\a_*)=\frac{1}{d-2}\frac{2}{S_d\b^{d-2}}Cl_{d-2}(\b\a_*+\pi)
\ee
In (\ref{idNielsen}) we substituted the Clausen function with odd index for the real part of the corresponding polylogarithm.

The calculation of the fermionic charge gap equations requires the sum
\be
\label{Qsum1}
\lim_{\e\rightarrow 0}\sum_{n=-\infty}^\infty\frac{e^{i\omega_n\epsilon}(\omega_n-\alpha_*)}{\bp^2+(\omega_n-\alpha_*)^2+\sigma_*^2}\,.
\ee
Without the introduction of the convergence factor $e^{i\omega_n\epsilon}$, $\e>0$ the sum would be undetermined \cite{SilvaNeto:1998dk}. Doing then the integral in the rhs of (\ref{Poisson}) term by term we first note that the $n=0$ term vanishes and we obtain (recall $\omega_n=\pi(2n+1)/\b$ here)
\be
\label{Qsumfer}
\lim_{\e\rightarrow 0}\sum_{n=-\infty}^\infty\frac{e^{i\omega_n\epsilon}(\omega_n-\alpha_*)}{\bp^2+(\omega_n-\alpha_*)^2+\sigma_*^2} = i\frac{\b}{2}\left(\frac{1}{1+e^{\b\sqrt{\bp^2+\s^2_*}+i\b\a_*}}-\frac{1}{1+e^{\b\sqrt{\bp^2+\s^2_*}-i\b\a_*}}\right)\,.
\ee
The corresponding bosonic sum with $\omega_n=2n\pi/b$ gives
\be
\label{Qsumbos}
\lim_{\e\rightarrow 0}\sum_{n=-\infty}^\infty\frac{e^{i\omega_n\epsilon}(\omega_n-\alpha_*)}{\bp^2+(\omega_n-\alpha_*)^2+\sigma_*^2} = i\frac{\b}{2}\left(\frac{1}{1-e^{\b\sqrt{\bp^2+\s^2_*}+i\b\a_*}}-\frac{1}{1-e^{\b\sqrt{\bp^2+\s^2_*}-i\b\a_*}}\right)\,.
\ee
We then define
\be
\label{defjd}
j_d(\sigma_*,\a_*)=\int\frac{d^{d-1}\bp}{(2\pi)^{d-1}}\frac{1}{1+e^{\b\sqrt{\bp^2+\s^2_*}+i\b\a_*}}
\ee
Notice that the cutoff dependence has dropped out i.e. the charge gap equation is finite. The same change of variables as in (\ref{id}) leads to
\be
\label{jd}
j_d(\sigma_*,\a_*)=-\frac{S_{d-1}}{(2\pi)^{d-1}}\frac{1}{\b^{d-1}}\int_0^{z_*}\frac{dz}{z+1}\left(\ln z+i\b\a_*\right)\left[(\ln z+i\b\a_*)^2-\b^2\s_*^2\right]^{\frac{d-3}{2}}
\ee
which enable us to calculate
\be
\label{Imjd}
\Im[j_{d}(\sigma_*,\a_*)]=\frac{1}{2i}[j_{d,f}(\sigma_*,\a_*)-\bar{j}_{d,f}(\sigma_*,\a_*)]
\ee
as a sum of $D_d(z)$ functions with even index. For $\s_*=0$ the integral (\ref{jd}) can also be calculated using Nielsen's polylogarithms and we obtain
\be
\label{jdNielsen}
j_d(0,\a_*)=i\frac{2}{S_d}\frac{1}{\b^{d-1}}Cl_{d-1}(\b\a_*+\pi)
\ee

From the result above we can write the expressions for the fermionic and bosonic chagres as
\begin{align}
\label{Qdfer}
Q_d&=i({\rm Tr}{\mathbb I}_{d-1})\Im[j_d(\sigma_*,\a_*)]\\
\label{qdbos}
q_d&=-2i\Im[j_d(\sigma_*,\a_*+\frac{\pi}{\b})]
\end{align}
Clearly, under $\b\a_*\leftrightarrow \b\a_*+\pi$ we have $2Q_d/{\rm Tr}{\mathbb I}_{d-1}\leftrightarrow -q_d$. Finally, the corresponding expressions for the charges at $\s_*=0$ are
\begin{align}
\label{Qdfer0}
Q_d\Bigl|_{\s_*=0}&=i{\rm Tr}{\mathbb I}_{d-1}\frac{2}{S_d\b^{d-1}}Cl_{d-1}(\b\a_*+\pi)\\
\label{qdbos0}
q_d\Bigl|_{\s_*=0}&=-2i\frac{2}{S_d\b^{d-1}}Cl_{d-1}(\b\a_*)
\end{align}

To evaluate the free energy density of the GN model in $d=5$ we use the following results
\begin{align}
\label{Res1}
&\frac{1}{\b}\int\frac{d^{4}\bar{p}}{(2\pi)^4}\Re\left[\ln\left(1+e^{-\b\sqrt{\bar{p}^2+\s_*^2}-i\b\a_*}\right)\right]=\nonumber \\
&\hspace{1cm}=\frac{3}{4\pi^2\b^5}\left[D_5(-z_*)+\frac{1}{6}\ln^2|z_*|D_3(-z_*)+\frac{1}{24}\ln^4|z_*|D_1(-z_*)+\frac{1}{90}\ln^5|z_*|\right]\\
\label{Res2}
&\frac{1}{2}\int\frac{d^5p}{(2\pi)^5}\ln\left(\frac{p^2+s_*^2}{p^2}\right)=\frac{\s_*^5}{120\pi^2}\\
\label{Res3}
&\frac{1}{2}\int^\Lambda\!\!\!\frac{d^5p}{(2\pi)^5}\frac{\s_*^2}{p^2+\s_*^2}=\frac{\s_{*}^2}{2G_{5,*}}+\frac{\s_*^5}{48\pi^2}-\frac{\s_*^4\Lambda}{24\pi^3} {}_2F_1\left(1,-\frac{1}{2};\frac{1}{2};-\frac{\s_*^2}{\Lambda^2}\right)\\
\label{Res4}
&\Re[i_5(\s_*,\a_*)]=-\frac{1}{4\pi^2\b^5}\left[D_3(-z_*)+\frac{1}{2}\ln^2|z_*|D_1(-z_*)+\frac{1}{6}\ln^3|z_*|\right]\\
\label{Res5}
&\Im[j_4(\s_*,\a_*)]=-\frac{1}{8\pi^2\b^4}\left[D_4(-z_*)+\frac{1}{6}D_2(-z_*)\right]
\end{align}

\section{The Bloch-Wigner-Ramakrishnan functions $D_d(z)$}
From the usual analytic continuation of the polylogarithms
\be
Li_d(z)=\sum_{n=1}^\infty\frac{z^n}{n^d}\,,\,\,\,z\in{\mathbb C}\setminus [1,\infty)\,,\,\,\,d=1,2,3,..\,.
\ee
one can define the following Bloch-Wigner-Ramakrishnan functions \cite{Zagier1,Zagier2} as
\be
\label{Ds}
D_d(z)=\Re\left(i^{d+1}\left[\sum_{k=1}^d\frac{(-\ln|z|)^{d-k}}{(d-k)!}Li_k(z)-\frac{(-\ln|z|)^d}{2d!}\right]\right)
\ee
These are real functions of complex variable, analytic in ${\mathbb C}\setminus \{0,1\}$. The functions $D_1(z)$ and $D_2(z)$ - the latter being the original Bloch-Wigner function - are given by
\be
\label{D12}
D_1(z)=\Re[\ln(1-z)]-\frac{1}{2}\ln|z|\,,\,\,\,\,\,\,\,\,
D_2(z)=\Im[Li_2(z)]+\ln|z|{\rm Arg}(1-z)
\ee
In the text we used the following properties of $D_d(z)$'s.
\begin{align}
\label{Dd1}
D_d(1/z)&=(-1)^{d-1}D_d(z)\\
\label{Dd2}
 \frac{\partial}{\partial z}D_d(z)&=\frac{i}{2z}\left(D_{d-1}(z)+\frac{i}{2}\frac{(-i\ln|z|)^{d-1}}{(d-1)!}\frac{1+z}{1-z}\right)
 \end{align}
On the unit circle  we have
\begin{align}
\label{Dodd}
&D_{2n-1}(e^{-i\theta})=(-1)^n\Re[Li_{2n-1}(e^{-i\theta})]=(-1)^n Cl_{2n-1}(\theta)\,,\\
\label{Deven}
&D_{2n}(e^{-i\theta})=(-1)^{n+1}\Im[Li_{2n}(e^{-i\theta})]=(-1)^{n}Cl_{2n}(\theta)\,
\end{align}
for $n=1,2,3,..$. 
The Clausen functions $Cl_m(\theta)$ are defined as
\be
\label{Clausen2}
Cl_{2n-1}(\theta)\equiv\sum_{k=1}^{\infty}\frac{\cos k\theta}{k^{2n-1}}\,,\,\,\,Cl_{2n}(\theta)\equiv \sum_{k=1}^{\infty}\frac{\sin k\theta}{k^{2n}}\,,\,\,\,n=1,2,..
\ee
and hence they are respectively even/odd functions of $\theta$. For example
\begin{align}
\label{Clausen1}
&D_1(e^{-i\theta})=Cl_1(-\theta)=-\ln|2\sin(\theta/2)|\,,\,\,\,D_2(e^{-i\theta})=Cl_2(-\theta)=-Cl_2(\theta)\\
&D_{2n-1}(e^{i\theta})=Cl_{2n-1}(-\theta)=Cl_{2n-1}(\theta)\,,\,\,\,D_{2n}(e^{-i\theta})=Cl_{2n}(-\theta)=-Cl_{2n}(\theta)\,.
\end{align}

\section{Thermal free energy of free fermions and scalars in $d$-dimensions}

For free massive $N$ Dirac fermions $\psi^a$, $\bar{\psi}^a$, $a=1,2,..,N$ in  $d$ Euclidean dimensions we have
\be
\label{diracaction}
{\cal I}_f=-\int ^{\b}_0\!\!dx^0 \!\!\int \!\!d^{d-1}\bar{x}\,(\bar{\psi}^a\gamma^\m\partial_\m\psi^a +m_f\bar{\psi}^a\psi^a)\,.
\ee
We put the theory in Euclidean $S_1\times\mathbb{R}^{d-1}$ where $S_1$ has radius $L=\beta=1/T$, $x^{0}\in[0,\b]$ and impose antiperiodic boundary conditions on the thermal circle resulting in
\be
\label{omegabos}
p_\m=(\omega_n,p^i)\,,\,\,\bar{p}=(p^i)\,,\,\,\omega_n=\frac{\pi}{\b}(2n+1)\,,\,\,\,n=0,\pm1,\pm2,..\,.
\ee
The thermal free energy  density $f_f(\b)$ is  defined as
\be
\label{Zf}
Z_f=\int({\cal D}\psi^a)({\cal D}\bar{\psi}^a)e^{-{\cal I}{_f}}
\equiv e^{-\beta V_{d-1}f_f^{free}(\b)}\,,
\ee
with $V_{d-1}$ the (infinite) volume of $\mathbb{R}^{d-1}$. The interesting quantity is the difference $f_f(\infty)-f_f(\b)\equiv \Delta f_f(\b)$ 
which is expected to be positive in a stable theory since $f_f(\b)=-{\cal P}_f(\b)$ is the fermionic pressure density at temperature $T=1/\b$. We obtain 
\begin{align}
\label{ffcalc}
\frac{1}{N\Tr\mathbb{I}_{d-1}}\D f^{(d)}_{f,free}(\b)&= \frac{1}{2}\int\frac{d^dp}{(2\pi)^d}\ln\left(\frac{p^2+m_f^2}{p^2}\right)+\frac{1}{\b}\int\frac{d^{d-1}\bar{p}}{(2\pi)^{d-1}}\ln\left(1+e^{-\b\sqrt{\bp^2+m_f^2}}\right)
\nonumber \\
&=\frac{\pi S_d}{2d(2\pi)^d}\left[\frac{m_f^d}{\sin\frac{\pi d}{2}}-\frac{S_{d-1}}{S_{d}}\frac{4d}{\b^d}\int_0^{e^{-\b m_f}}\left(\ln^2x-m_f^2\b^2\right)^{\frac{d-3}{2}}\ln x\ln(1+x)\frac{dx}{x}\right]\,.
\end{align}
For $m_f=0$ we obtain
\be
\label{Dfd}
\frac{1}{N\Tr\mathbb{I}_{d-1}}\D f^{(d)}_{f,free}(\b)=-\frac{2}{\b^d S_d}Li_d(-1)=(-1)^{\frac{d-1}{2}}\frac{2}{\b^d S_d}D_d(-1)=-\frac{2}{\b^d S_d}\left(1-\frac{1}{2^{d-1}}\right)\zeta(d)
\ee
For $d=1$ we can use $Li_1(-1)=-\ln 2$ to find the the free energy of a massless 1$d$ fermion is $T\ln 2$.


For real scalars $\phi^a(x)$ with Euclidean action 
\be
\label{scalaraction}
{\cal I}_b=\int ^{L}_0\!\!dx^0 \!\!\int \!\!d^{d-1}\bar{x}\left(\frac{1}{2}\partial_\m\phi^a\partial_\m\phi^a +\frac{1}{2}m_b^2\phi^a\phi^a\right)\,.
\ee
we obtain, using periodic boundary conditions on the thermal circle and the corresponding bosonic frequencies $\omega_{n}=2\pi/\beta$,
\begin{align}
\label{fbcalc}
\frac{1}{N}\Delta f^{(d)}_{b,free}(\b) &= -\frac{1}{2}\int\frac{d^dp}{(2\pi)^d}\ln\left(\frac{p^2+m_b^2}{p^2}\right)-\frac{1}{\b}\int\frac{d^{d-1}\bar{p}}{(2\pi)^{d-1}}\ln\left(1-e^{-\b\sqrt{\vec{p}^2+m_b^2}}\right)
\nonumber \\
 &=-\frac{\pi S_d}{4d(2\pi)^d}\left[\frac{m_b^d}{\sin\frac{\pi d}{2}}-4d\frac{S_{d-1}}{S_d}\frac{1}{\b^d}\int_0^{e^{-\b m_b}}\left(\ln^2x-m_b^2\b^2\right)^{\frac{d-3}{2}}\ln x\ln(1-x)\frac{dx}{x}\right]\,,
 \end{align}
 For $m_b=0$ we obtain
 \be
\label{Dbd}
\frac{1}{N}\D f^{(d)}_{b,free}(\b)=\frac{2}{\b^d S_d}Li_d(1)=\frac{2}{\b^d S_d}D_d(-1)=\frac{2}{\b^d S_d}\zeta(d)
\ee

\bibliographystyle{JHEP}
\bibliography{Refs}

\end{document}